\documentclass[submission, Phys]{SciPost}

\usepackage{amsfonts}
\usepackage{amsmath}
\usepackage{physics}
\usepackage{graphicx}
    \usepackage{subcaption}
\usepackage{manfnt}
\usepackage{tikz}
    \usepackage{tikzscale}
    \usepackage{xifthen}
    \usetikzlibrary{decorations.markings}
    \newcommand{\tikzXe}{\begin{tikzpicture}[baseline={([yshift=-.5ex]current bounding box.center)}]
        \draw[draw=gray](0,0)to(1,0) node[text=red!50] at (.5,0) {$X$};
    \end{tikzpicture}}

\newcommand{\tikzAv}{\begin{tikzpicture}[baseline={([yshift=-.5ex]current bounding box.center)}]
        \draw[draw=gray](0,0) node[text=red!50] at(.5,0) {$Z$} to (2,0) node[text=red!50] at(1.5,0) {$Z$};
        \draw[draw=gray](1,-1) node[text=red!50] at (1,-.5) {$Z$} to (1,1) node[text=red!50] at (1,.5) {$Z$} 
        ;
    \end{tikzpicture}}

\newcommand{\tikzXv}{\begin{tikzpicture}[baseline={([yshift=-.5ex]current bounding box.center)}]
        \draw[draw=gray](0,0) to (1,0);
        \draw[draw=gray](.5,-.5) node[text=blue!50] at (.5,0) {$X$} to (.5,.5)
        ;
    \end{tikzpicture}}

\newcommand{\tikzZZ}{\begin{tikzpicture}[baseline={([yshift=-.5ex]current bounding box.center)}]
        \draw[draw=gray](0,0) node[text=blue!50]{$Z$} to (1,0) node[text=blue!50] {$Z$};
    \end{tikzpicture}}

\newcommand{\tikzZZonC}{\begin{tikzpicture}[baseline={([yshift=-.5ex]current bounding box.center)},decoration={markings, mark= at position 0.35 with {\arrow[orange!70, line width=2pt]{>}}}]
        \draw[draw=gray](0,0) node[text=blue!50]{$Z$} to (0,1) node[text=blue!50] {$Z$};
        \draw[draw=gray](-.5,0) to (+.5,0);
        \draw[draw=gray](-.5,1) to (.5,1);
        \draw[line width = 3pt, draw=orange!50, postaction={decorate}] (-.25,1)to(-.25,0);
        \draw[line width = 3pt, draw=orange!50, postaction={decorate}](+.25,0) to (+.25,1);
    \end{tikzpicture}}

\newcommand{\tikzZXZonC}{\begin{tikzpicture}[baseline={([yshift=-.5ex]current bounding box.center)},decoration={markings, mark= at position 0.35 with {\arrow[orange!70, line width=2pt]{>}}}]
        \draw[draw=gray](0,0) node[text=red!50]{$Z$} node at (.5,0)[text=blue!50] {$X$} to (1,0) node[text=red!50] {$Z$};
        \draw[draw=gray](.5,-.5) to (.5,.5);
        \draw[line width = 3pt, draw=orange!50, postaction={decorate}](.25,.5) to (.25,-.5);
        \draw[line width = 3pt, draw=orange!50, postaction={decorate}](.75,-.5) to (.75,.5);
    \end{tikzpicture}}

\newcommand{\tikzZX}{\begin{tikzpicture}[baseline={([yshift=-.5ex]current bounding box.center)}]
        \draw[draw=gray](0,0) node[text=blue!50]{$Z$} to (1,0) node[text=red!50] at (.5,0){$X$} ;
        \draw[line width = 3pt, draw=orange!50](.25,-.5) to (.25,.5) node[text=orange] at (.25,-.8) {$\mathcal{W}$};
    \end{tikzpicture}}

\newcommand{\tikzCondConstraint}{\begin{tikzpicture}[baseline={([yshift=-.5ex]current bounding box.center)},decoration={markings, mark= at position 0.53 with {\arrow[orange!70, line width=2pt]{>}}}]
    \draw[line width=1pt, draw=gray!50] (3.5,.5) grid (5.5,2.5);
    \node[text=orange!50] at (4.5,.5) {$\Sigma_1$};
    \draw[line width=3pt, draw=orange!50, postaction={decorate}] (4.25,.5) to (4.25,2.5);
    \draw[line width=3pt, draw=orange!50, postaction={decorate}] (4.75,2.5)to(4.75,.5);   
    \node[text=red!50] at (4.5,1){$X$};
    \node[text=red!50] at (4.5,2){$X$};
\end{tikzpicture}}

\newcommand{\tikzXXXX}{\begin{tikzpicture}[scale=.5,baseline={([yshift=-.5ex]current bounding box.center)}]
        \draw[draw=gray](0,0) node[text=red!50] at(1,0) {$X$} to (2,0);
        \draw[draw=gray](0,0) node[text=red!50] at (0,1) {$X$} to (0,2);
        \draw[draw=gray](2,0) node[text=red!50] at (2,1) {$X$} to (2,2);
        \draw[draw=gray](0,2) node[text=red!50] at (1,2) {$X$} to (2,2);
\end{tikzpicture}}

\newcommand{\tikzXXhor}{\begin{tikzpicture}[scale=.5,baseline={([yshift=-.5ex]current bounding box.center)}]
        \draw[draw=gray](0,0) to (2,0);
        \draw[draw=gray](0,0) node[text=red!50] at (0,1) {$X$} to (0,2);
        \draw[draw=gray](2,0) node[text=red!50] at (2,1) {$X$} to (2,2);
        \draw[draw=gray](0,2) to (2,2);
\end{tikzpicture}}

\newcommand{\tikzXsquare}{\begin{tikzpicture}[scale=.5,baseline={([yshift=-.5ex]current bounding box.center)}]
        \draw[draw=gray](0,0) to (2,0);
        \draw[draw=gray](0,0) to (0,2);
        \draw[draw=gray](2,0) node[text=red!50] at (2,1) {$X$} to (2,2);
        \draw[draw=gray](0,2) to (2,2);
\end{tikzpicture}}

\newcommand{\tikzXXX}{\begin{tikzpicture}[scale=0.5, baseline={([yshift=-.5ex]current bounding box.center)}]
        \draw[draw=gray](0,0) node[text=red!50] at(1,0) {$X$} to (2,0);
        \draw[draw=gray](0,0) node[text=red!50] at(0,1) {$X$} to (0,2);
        \draw[draw=gray](2,0) to (2,2);
        \draw[draw=gray](0,2) node[text=red!50] at (1,2) {$X$} to (2,2);
\end{tikzpicture}}

\usepackage{bm}
\usepackage{amssymb}
\usetikzlibrary{3d,calc}
\usepackage{mathtools}
\usepackage{enumitem}
\usetikzlibrary{calc}

\usepackage{color}

\newcommand{\cA}{\mathcal{A}}
\newcommand{\cB}{\mathcal{B}}
\newcommand{\cC}{\mathcal{C}}
\newcommand{\cD}{\mathcal{D}}
\newcommand{\cH}{\mathcal{H}}
\newcommand{\cU}{\mathcal{U}}
\newcommand{\cV}{\mathcal{V}}
\newcommand{\cW}{\mathcal{W}}
\newcommand{\C}{\mathbb{C}}
\newcommand{\Z}{{\mathbb Z}}

\begin{document}

\begin{center}{\Large \textbf{
Lattice Gauging Interfaces and Non-invertible Defects in Higher Dimensions
}}\end{center}

\begin{center}
David Hofmeier\textsuperscript{1,*},
Giovanna Pimenta \textsuperscript{1,2},
Weiguang Cao\textsuperscript{1,3}
\end{center}

\begin{center}
{\bf 1} Center for Quantum Mathematics at IMADA, Southern Denmark University, Campusvej 55, 5230 Odense, Denmark
\\
{\bf 2} Department of Physics, State University of Londrina, Londrina, PR, Brazil 
\\
{\bf 3} Niels Bohr International Academy, Niels Bohr Institute, University of Copenhagen, Blegdamsvej 17, DK-2100 Copenhagen, Denmark\\
*hofmeier@imada.sdu.dk
\end{center}

\begin{center}
\today
\end{center}


\section*{Abstract}
{\bf
We study gauging interfaces and their defect descendants in lattice models with generalized symmetries in higher dimensions. We construct explicit interface Hamiltonians for gauging a $\mathbb Z_2^{(0)}$ symmetry in $(2+1)d$ and a $\mathbb Z_2^{(1)}$ symmetry in $(3+1)d$. In higher dimensions, and especially in the presence of higher-form symmetries, the topological nature of gauging interfaces is obscured by the fact that the constrained Hilbert space depends on the location of the interface. We resolve this by introducing movement operators acting on a common unconstrained Hilbert space, which transport both the interface Hamiltonians and the associated constraints.
As applications, we analyze condensation defects obtained from finite-region gauging and reconstruct the gauging map from movement operators. Finally, we apply the same framework to subgroup gauging, focusing on the example of gauging $\mathbb Z_2\subset \mathbb Z_4$. This produces a dual symmetry carrying a mixed anomaly, which we diagnose on the lattice through symmetry fractionalization on condensation defects.
Our results provide an explicit lattice framework for studying topological interfaces, condensation defects, and the associated anomalies arising from gauging in higher-dimensional systems.
}

\vspace{10pt}
\noindent\rule{\textwidth}{1pt}
\tableofcontents
\noindent\rule{\textwidth}{1pt}
\vspace{10pt}

\section{Introduction}
\label{sec:intro}
Global symmetry and its modern generalizations~\cite{gaiotto2015generalized}, such as higher-form symmetries~\cite{Sharpe:2015mja,Hsin:2018vcg,Cordova:2018cvg,Lake:2018dqm,Benini:2018reh,Harlow:2018tng,Eckhard:2019jgg} and non-invertible symmetries~\cite{Bhardwaj2018,Tachikawa:2017gyf,Thorngren:2019iar,Thorngren:2021yso,Bhardwaj:2022yxj,Bhardwaj:2022lsg,Bhardwaj:2022kot}, play important roles in many areas of quantum field theory~\cite{Choi:2022jqy,Cordova:2022ieu,Cordova:2022fhg,Cordova:2024ypu} and condensed matter physics~\cite{Li:2023knf,Bhardwaj:2023idu,Bhardwaj:2023fca,Bhardwaj:2023bbf,Bhardwaj:2024qrf,Bhardwaj:2024wlr,Bhardwaj:2024qiv,Warman:2024lir,Li:2024gwx,Inamura:2025cum,Li:2026aja}. A basic feature of these generalized symmetries is their topological nature~\cite{Chang:2018iay}: when a symmetry operator is inserted in a correlation function, its supporting manifold can be deformed without affecting the result, as long as the deformation does not cross charged operators. In continuum field theories, this makes the symmetry operator (where the supporting manifold is fully spatial) and the symmetry defect (where the supporting manifold has one component in the time direction) treated on equal footing.

On the lattice, however, this topological picture is more subtle~\cite{Aasen2016}. Since space and time are treated asymmetrically, the equivalence between symmetry operators and symmetry defects, as well as the topological property of symmetry defects in the microscopic Hilbert space~\cite{Aasen:2020jwb,Seifnashri:2023dpa}, is no longer automatic. This problem has attracted increasing attention in recent years, especially in the study of lattice realizations of Kramers--Wannier (KW) duality~\cite{Seiberg:2023cdc,Seiberg2024} and many other non-invertible structures~\cite{Lootens:2021tet,Lootens:2022avn,Lu:2024ytl}. In particular, many interesting non-invertible symmetries in lattice models arise from gauging~\cite{Li:2023ani,Cao:2023doz,Yan:2024eqz,Choi:2024rjm,Gorantla:2024ocs,Seifnashri:2024dsd,Seifnashri:2025fgd,Cao:2025qhg,Cao:2025xep,Ortiz2026}, or from compositions of gauging transformations with invertible operations~\cite{Cao:2024qjj,Pace:2024oys,Hsin:2024aqb,Cao:2025qnc,Hsin:2025ria}. It is therefore natural to regard gauging interfaces as elementary building blocks for a broader class of topological defects with non-invertible structures on the lattice.

The goal of this paper is to develop an explicit lattice framework for such gauging interfaces in higher dimensions~\cite{Choi:2021kmx,Kaidi:2021xfk}. The main subtlety is that in higher-dimensional lattice systems and especially in the presence of higher-form symmetries the topological nature of the interface is obscured by the fact that moving the interface changes the constrained Hilbert space~\cite{Tantivasadakarn:2023zov}. Indeed, different interface locations generally correspond to different identifications of local degrees of freedom and to different no-twist or zero-flux constraints. Therefore, making the topological character of the interface explicit requires more than writing down interface Hamiltonians at fixed positions: one must also construct movement operators that transport both the Hamiltonian and the associated constraints.

In this paper, we address this problem through two basic lattice examples. We study the lattice gauging interface obtained by gauging a $\mathbb Z_2^{(0)}$ symmetry in $(2+1)d$ and gauging a $\mathbb Z_2^{(1)}$ symmetry in $(3+1)d$~\cite{Koide:2021zxj}. By gauging on only half of the space, we construct the corresponding interface Hamiltonian separating the original theory from its gauged dual. To show the topological nature of the gauging interface~\cite{Tantivasadakarn:2025txn}, we explicitly formulate the interface on a common unconstrained Hilbert space and construct explicit local movement operators that move the interface while transforming the Hamiltonian and the constraints simultaneously. 

As one application, we analyze condensation defects~\cite{Roumpedakis2023,Choi2023,Cordova:2024mqg} obtained by gauging a connected finite sublattice. In the $(2+1)d$ and $(3+1)d$ examples, we study their fusion with invertible symmetry defects, their self-fusion, and the relation between contractible and non-contractible condensation defects under movement and projection. In the $(2+1)d$ case, this viewpoint also reconstructs the gauging map from a decomposable condensation defect by sweeping the associated interface across the full spatial lattice.

Taken together, these examples can be straightforwardly generalized to arbitrary Abelian groups and spacetime dimensions, and thus provide a concrete lattice framework for studying topological interfaces and their defect descendants in higher-dimensional systems. Finally, we apply this framework to subgroup gauging for $\mathbb Z_2 \subset \mathbb Z_4$, where gauging produces a partially gauged model with dual residual symmetries carrying a mixed anomaly~\cite{Bhardwaj2023}. We show how the resulting anomaly is diagnosed on the lattice through defect fusion and symmetry fractionalization on condensation defects~\cite{Moradi2025}. This gives another concrete application of our framework to a more general anomalous symmetry structure.

The paper is organized as follows. In Sec.~\ref{sec:0form}, we study gauging interfaces arising from gauging a 
$\mathbb Z_2^{(0)}$ symmetry in $(2+1)d$. We begin by reviewing the lattice gauging map, then construct the corresponding interface Hamiltonians and the fusion between gauging interface with symmetry defects, and finally develop the movement operators on a common unconstrained Hilbert space. In Sec.~\ref{sec:0formCond}, we apply this framework to condensation defects in the same $(2+1)d$ setting, including both contractible and non-contractible defects and their fusion properties. In Sec.~\ref{sec:1form}, we turn to gauging a $\mathbb Z_2^{(1)}$ symmetry in $(3+1)d$, construct the corresponding gauging interfaces, and analyze the associated condensation defects. In Sec.~\ref{sec:Subsym}, we consider gauging the subgroup $\mathbb Z_2 \subset \mathbb Z_4$, derive the corresponding sequential gauging relations, and use defects to diagnose the mixed anomaly of the partially gauged theory. We conclude in Sec.~\ref{sec:conclusion} with a discussion of the main results and possible future directions.

Throughout the paper, we denote the two orientations of the gauging interface by 
$\mathcal{D}$ and $\mathcal{D}^{\dagger}$. We write $\eta_p$ for the invertible defect associated with the $\mathbb{Z}_2^{(p)}$symmetry, and $\mathcal{C}_p$ for the corresponding condensation defect. We use $\cong$ to denote unitary equivalence of Hamiltonians or defect/interface configurations. Since gauging interfaces are topological and closely related to duality transformations, we will use the terms gauging interface, duality interface, and topological interface interchangeably when no confusion can arise.

\section{Gauging Interfaces from Gauging $\Z_2^{(0)}$ Symmetry in $(2+1)d$}
\label{sec:0form}
In this section, we construct lattice gauging maps and the associated topological gauging interfaces that arise from gauging a global $\Z_2^{(0)}$ symmetry in $(2+1)d$ in detail.
Our presentation will be straightforward for generalization to arbitrary dimension.
For example in $(1+1)d$, our construction reduces to the ordinary Kramers-Wannier (KW) duality map, interfaces and defects discussed in~\cite{Seiberg2024}.
In higher dimensions, gauging the $0$-form symmetry produces the quantum  $(d{-}1)$-form symmetry in the dual theory~\cite{Vafa:1989ih}. 

\subsection{Gauging on the Lattice}
\label{subsec:2dgauging}
Let us begin by reviewing gauging $\mathbb Z_2^{(0)}$ symmetry and the corresponding gauging map on a square lattice $\Gamma$ in two spatial dimensions. 
Our analysis is general and can be readily generalized to arbitrary Abelian groups and lattice configurations. 

\paragraph{$\Z_2^{(0)}$ symmetric lattice models.} 
To each vertex $v$ of $\Gamma$, we associate the state space $\mathbb{C}^{2}$, so that the total Hilbert space is 
\(
\mathcal{H}=\bigotimes_{v}\mathbb{C}^2_v
\).
We denote the Pauli operators on vertex $v$ by $X_v$ and $Z_v$.
The global $\Z_2^{(0)}$ symmetry is generated by
\begin{equation}
    U = \prod_{v\in\Gamma}X_v\,.
\end{equation}
Any local operator invariant under this symmetry can be generated by products of $X_v$, and  $Z_{s(e)}Z_{t(e)}$ along oriented edges $e$ of $\Gamma$, where $s(e)$ and $t(e)$ denote source and target vertices. 
Thus, the space of $\Z_2^{(0)}$ symmetric Hamiltonians is generated by the bond algebra built from $X_v$ and $Z_{s(e)}Z_{t(e)}$.
A simple example is the transverse-field Ising model (TFIM)
\begin{equation}
    H^{(0)} = -\sum_{e}Z_{s(e)}Z_{t(e)}-g\sum_vX_v\,.
    \label{eq:TFIM}
\end{equation}
It will also be convenient to define the operator
\begin{equation}
    V_\gamma = \prod_{e\in \gamma}Z_{s(e)}Z_{t(e)}\,,
\end{equation}
for any closed path $\gamma$ in the lattice. It measures the $\Z_2^{(0)}$ twist around $\gamma$. Its eigenvalues $\pm 1$ label the twist sectors and indicate whether $\gamma$ crosses an even or odd number of $\Z_2$-twist lines. 
To ensure that the dual symmetry after gauging is topological, we require $V_\gamma = 1$ for all contractible loops.
In what follows, these operators will serve as book-keeping devices that track how twisted sectors are mapped under gauging.

\paragraph{Gauging the $\Z_2^{(0)}$ symmetry.} 
We introduce $\mathbb Z_2$ gauge degrees of freedom on the edges $e$ and consider the extended Hilbert space \begin{equation}
    \cH_{\text{ext}} = \bigotimes_{v\in \Gamma}\C^2_v\bigotimes_{e\in \Gamma}\C^2_e,
\end{equation}
and $X_e,Z_e$ are Pauli operators acting on the gauge qubits on edge $e$.
Minimal coupling to the gauge field on the edge promotes the global $\Z_2^{(0)}$ invariant operators to gauge-invariant operators
\begin{equation}
    \{X_v\,, Z_{s(e)}Z_{t(e)}\,, V_{\gamma}\}\longmapsto \{X_v\,, Z_{s(e)}X_eZ_{t(e)}\,, V_{\gamma}\prod_{e\in \gamma}X_e\}\,.
    \label{eq:Z2 0fs gauged bond algebra}
\end{equation}
Gauge transformations are generated by local Gauß operators
\begin{equation}
    G_v = X_vA_v, \quad A_v=\prod_{e\ni v}Z_e, \quad \forall  v\in \Gamma,
\end{equation}
and physical states obey  $G_v=1$ for all $v$. 
We will use the unitary
\begin{equation}
   \mathcal{M} = \prod_v\prod_{e\ni v}\mathsf{CZ}_{e,v}:\quad 
    X_v\mapsto X_vA_v,\quad 
    X_e\mapsto Z_{s(e)}X_eZ_{t(e)},\quad
    Z_v \mapsto Z_v,\quad  
    Z_e\mapsto Z_e\,.
\end{equation}
to localize the Gauß constraints $\mathcal{M}G_v \mathcal{M}^{\dagger}=X_v=1$ such that we can project onto gauge invariant sectors with degrees of freedom purely on edges.

On this physical subspace, the original $\Z_2^{(0)}$ symmetry and twist generators are mapped to the gauged ones
\begin{equation}
\label{eq:gauging map}
    \{X_v\,, Z_{s(e)}Z_{t(e)}\,, V_{\gamma}\}\longmapsto \{A_v\,, X_e\,, \prod_{e\in \gamma}X_e\}\,.
\end{equation}
Note that the gauged local operators commute with the Wilson loops of the $\Z_2$ gauge field
\begin{equation}
    W_{\gamma}=\prod_{e\in \gamma}X_e\,,
\end{equation}
which generate the quantum $\Z_2^{(1)}$ symmetry.
Under gauging, the twist-measuring operators $V_\gamma$ are mapped to these symmetry operators. The zero-flux constraint $W_{\partial p} = 1$ for all contractible loops $\partial p$ implies the gauged Hilbert space is non-factorizable, as usual in lattice gauge theory.

Applying this gauging procedure to the TFIM \eqref{eq:TFIM} yields the Hamiltonian of $\mathbb Z_2$ lattice gauge theory
\begin{align}
    \begin{split}
    H^{(1)} &= -\sum_e X_e-g\sum_vA_v=-\sum_e\tikzXe
    -g\sum_v\tikzAv\,.
    \label{eq:2dLGT}
    \end{split}
\end{align}
Conversely, starting from~\eqref{eq:2dLGT}, gauging the $\Z_2^{(1)}$ symmetry returns the $\Z_2^{(0)}$ symmetric TFIM Hamiltonian~\eqref{eq:TFIM}.
Thus, these two Hamiltonians are dual to each other.

\paragraph{Gauging map.}
It is conceptually useful to package the above gauging procedure into a single linear map
 \begin{equation}
     \mathsf{D}=  \otimes_v \bra{+}_v\, \mathcal{M}\,\otimes_e \ket{+}_e\,,
     \label{eq:2dGaugingMap}
 \end{equation}
where we first couple the theory to edge gauge field through adding the product state $\otimes_e \ket{+}_e$, then apply the unitary $\mathcal{M}$ to localize the Gauß constraints, and finally project onto the gauge invariant sector with $\otimes_v \bra{+}_v$.
Its Hermitian conjugate $\mathsf{D}^{\dagger}$ implements gauging $\mathbb Z_2^{(1)}$ symmetry. 
It can be checked straightforwardly that $D$ acts on operators as
\begin{equation}
    \mathsf{D}X_v = A_v\mathsf{D} \,,\quad \mathsf{D}Z_{s(e)}Z_{t(e)} = X_e\mathsf{D}\,.
\end{equation}
The gauging map $\mathsf{D}$ does not have an inverse. Instead, it fulfills the non-invertible fusion rule \cite{Choi2023}
\begin{equation}\label{eq:fusionop2d}
    \mathsf{D}^{\dagger}\times \mathsf{D}=1+U,\quad \mathsf{D}\times\mathsf{D}^{\dagger}=(1+U_{\Sigma_x})(1+U_{\Sigma_y})\,,
\end{equation}  
which gives the projection operator for $\mathbb Z_2^{(0)}$ symmetry and the condensation operator for $\mathbb Z_2^{(1)}$ symmetry. 
The fusion of the gauging map with invertible symmetry operators is
\begin{equation}\label{eq:fusionwithinv1d}
\mathsf{D}U=W_{\gamma}\mathsf{D}=\mathsf{D},\quad U\mathsf{D}^{\dagger}=\mathsf{D}^{\dagger}W_{\gamma}=\mathsf{D}^{\dagger}\,.
\end{equation}
These fusion rules will be re-derived later using gauging interfaces.

\subsection{Gauging Interfaces}
\label{subsec:2dHalfGauging}
We now construct \emph{gauging interfaces} on the two-dimensional square lattice.
A gauging interface is a codimension-one submanifold whose worldvolume extends along time and spatially separates two regions. In our example, on one side is the $\Z_2^{(0)}$ symmetry, while on the other side this symmetry has been gauged to a $\Z_2^{(1)}$ symmetry.
On the lattice, we can construct an \emph{interface Hamiltonian} where the degrees of freedom and local terms differ on the two sides of the interface but are glued consistently along the interface locus. \cite{Hauru2016,Aasen2016}

Because gauging can be implemented sequentially by local unitaries~\cite{Seiberg2024,Tantivasadakarn:2025txn}, the resulting interface is topological. On the lattice, this means that interfaces at different positions are unitarily related. Here, we construct these interfaces for $\Z_2^{(0)}$ symmetry and describe their movement and fusion, thus generalizing the $(1+1)d$ results in~\cite{Seiberg2024}.

\begin{figure}[h]
    \centering
    \includegraphics[width=0.75\linewidth]{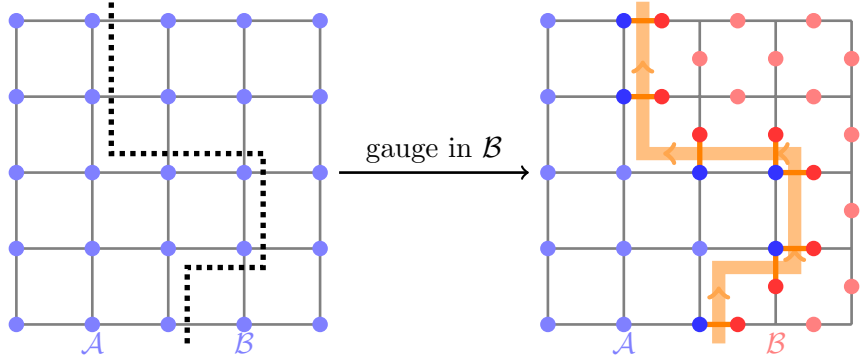}
    \caption{We gauge the semi-infinite region $\cB$ to localize a duality interface $\cD$ on $\cW$ that interpolates between a theory with global $\mathbb{Z}_2^{(0)}$ symmetry (blue, region $\cA$) and a theory with quantum $\mathbb{Z}_2^{(1)}$ symmetry (red, region $\cB$).
    The boundary sets $\cW_\cA$ and $\cW_\cB$ are indicated by darker shades.
    The arrow on $\cD$ denotes its orientation and distinguishes it from its conjugate $\cD^\dagger$.}
    \label{fig:2dDefect}
\end{figure}
\paragraph{Geometry of the interface.} We divide the lattice $\Gamma$ into two regions $\mathcal A$ and $\mathcal B$ such that $\Gamma=\cA\sqcup \cB$.
We gauge the $\Z_2^{(0)}$ symmetry only in region $\cB$, keeping region $\cA$ in the original description; see Fig.~\ref{fig:2dDefect}. At this point, we make no assumptions on the shape of the boundary. We do, however, always choose the two regions such that $\cA$ ends on vertices and $\cB$ ends on edges. The upside is that with such a choice, we do not have to alter the definition of the Gauß law close to the interface in $\cB$. After gauging, the interface is localized on the joint boundary
\begin{equation}
   \cW = \cW_\cA \sqcup \cW_\cB\,,
\end{equation}
where $\cW_\cA$ consists of the boundary vertices of $\Gamma_{\cA}$, those attached to edges crossing the interface, and $\cW_\cB$ consists of the boundary edges of $\Gamma_{\cB}$, those crossing the interface.

\begin{figure}[h]
    \centering
    \includegraphics[width=0.75\linewidth]{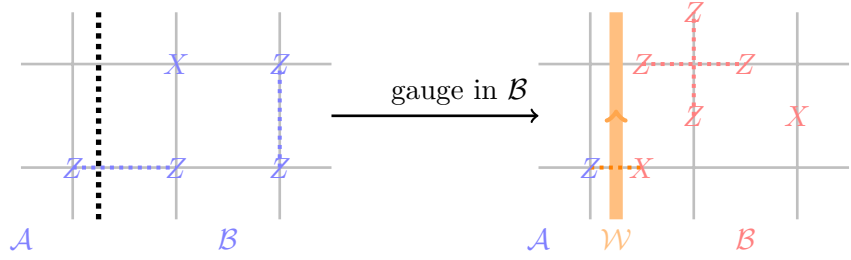}
    \caption{Local action of the gauging map in region $\cB$ in the presence of an interface.
    Edges in the bulk of $\cB$ are gauged as in the uniform case, while edges crossing the interface acquire a modified vertex–edge terms $Z_vX_e$.}
    \label{fig:GaugingMapOnLattice}
\end{figure}

\paragraph{Half space gauging and interface Hamiltonian.}
Gauging in region $\cB$ proceeds as before: we minimally couple the spins and apply $\mathcal M_v$ on all $v\in\cB$.
Inside $\cB$ (away from the interface), we have the usual gauging map and constraints
\begin{equation}
    X_v\to A_v,\qquad 
    Z_{s(e)}Z_{t(e)}\to X_e,\qquad 
    W_{\partial p}=1,
    \qquad v,p\in\cB,\; e\in\cB/\cW_\cB\,.
\end{equation}
For edges $e\in\cW_\cB$ crossing the interface, the gauging map is slightly modified because one endpoint $v\in e\cap\cW_\cA$ remains a vertex degree of freedom:
\begin{equation}
    Z_{s(e)}Z_{t(e)}\longmapsto Z_{v}X_e, 
    \qquad e\in \cW_{\cB},~ v\in e\cap\cW_{\cA}\,,
\end{equation}
which is illustrated in Figure \ref{fig:GaugingMapOnLattice}.
The plaquette constraints for plaquettes $p$ intersecting the interface also change to
\begin{equation}
    \prod_{e\in p\cap\cW_{\cB}}X_e=1,
    \qquad \text{if  }p\cap\cW\neq \emptyset\,.
\end{equation}

For example, starting from the TFIM~\eqref{eq:TFIM} and gauging only in region $\cB$, we obtain the interface Hamiltonian with a gauging interface $\mathcal{D}(\cW)$ located on $\cW$
\begin{equation}
     H_{\mathcal{D}(\cW)} = H_{\mathcal A}+H_{\mathcal B}+H_{\cW},
    \label{eq:D01Defect}
\end{equation}
where 
\begin{align}\label{eq: half gauging of Ising}
    \begin{split}
        H_{\mathcal A}&=-\sum_{e\in\cA}Z_{s(e)}Z_{t(e)} - \sum_{v\in\cA}X_v = -\sum_{e\in\cA}\tikzZZ - \sum_{v\in\cA}\tikzXv\,,\\
        H_{\cB}&=- \sum_{v\in\cB} A_v - \sum_{e\in\cB/\cW_{\cB}}X_e = - \sum_{v\in\cB}\tikzAv - \sum_{e\in\cB/\cW_{\cB}}\tikzXe\,,\\
        H_{\cW}&=- \sum_{e\in \cW_{\cB}, v\in e\cap\cW_{\cA}}Z_vX_e = -\sum_{e\in \cW_{\cB}, v\in e\cap\cW_{\cA}}\tikzZX\,,
    \end{split}
\end{align}
are the parts of the Hamiltonian in regions $\mathcal A$, $\mathcal B$ and localized around $\mathcal W$. Away from the interface, $H_{\cA}$ and $H_{\cB}$ reduce to the TFIM Hamiltonian~\eqref{eq:TFIM} and the $\Z_2$ lattice gauge theory~\eqref{eq:2dLGT}, and $H_{\cW}$ in~\eqref{eq:D01Defect} interpolates between the two theories.\footnote{In Ref.~\cite{Giridhar2025}, this half-gauging procedure has also been used to derive the defect Hamiltonian associated to the Xu-Moore model.} Note that in the above and throughout the following, we fix the coupling constant $g=1$ for ease of notation.
Analogously, gauging region $\cB$ in the $\Z_2$ gauge theory produces the conjugate interface $\mathcal{D}^\dagger(\cW)$.

One can also obtain a unitarily equivalent presentation of the same interface by starting from $\Z_2$ gauge theory and gauging $\Z_2^{(1)}$ in region $\cA$. This yields
\begin{equation}
    \widetilde{H}_{\mathcal{D}(\cW)} = \widetilde{H}_{\mathcal A}+\widetilde{H}_{\mathcal B}+\widetilde{H}_{\cW},
\end{equation}
with 
\begin{align}
    \begin{split}
        \widetilde{H}_{\mathcal A}&=-\sum_{e\in\cA}Z_{s(e)}Z_{t(e)}-\sum_{v\in\cA/\cW_\cA}X_v,\\
        \widetilde{H}_{\mathcal B}&=-\sum_{v\in\cB}A_v-\sum_{e\in\cB}X_e,\\
        \widetilde{H}_{\cW}&=-\sum_{v\in\cW_\cA}X_v\left(\prod_{e\ni v, e\in\cW_\cB}Z_e\right)\,.
    \end{split}
    \label{eq:0formAlternativeInterface}
\end{align}
The two presentations are related by a unitary localized on the interface
\begin{equation}
    \widetilde{H}_{\mathcal{D}(\cW)}=U_{\cW}\,H_{\mathcal{D}(\cW)}\,U_{\cW}^{\dagger},\quad U_{\cW}=\prod_{<v,e>\in\cW_\cA,\cW_\cB}\mathsf{CZ}_{v,e}.
    \label{eq:2dRelationBtwPresentations}
\end{equation}
Both presentations will be useful below. For example, it will turn out to be convenient to use opposite presentations for two neighboring interfaces when analyzing their fusion.

\paragraph{Fusions with symmetry defects.}
We now compute how the duality interface $\cD(\cW)$ fuses with symmetry defects of the 0-form and 1-form symmetries. We denote by $\eta_p$ the defect of $\mathbb Z_2^{(p)}$ symmetry. In general, the symmetry defects of ordinary, invertible onsite symmetries can always be obtained by acting with a truncated symmetry operator, with the defect then being located at its boundary. For $p=0$, $\eta_p$ is defined on a one-dimensional sublattice $\Sigma_1^\vee$, defined on the dual lattice. It is created by applying $\prod_vX_v$ to some sublattice. Because this defect creation operator is a product of local unitaries, the invertible symmetry defects can easily be moved by acting with local unitaries that change the region of symmetry action. Similarly, for $p=1$, $\eta_p$ is defined on a single vertex $\Sigma_0$ and it is created by acting with $\prod_eX_e$ along some 1d string that ends on $\Sigma_0$.

\medskip\noindent Let us now first consider a $\Z_2^{(0)}$ symmetry defect $\eta_0$. We create $\eta_0$ somewhere in $\cA$ and then move it to the interface $\cW$. This joint action is represented by the unitary $\prod_{v\in\cA}X_v$ on the Ising side, which flips the sign of $Z_{s(e)}Z_{t(e)}$ on edges crossing $\cW$:
\begin{equation}
-\sum_{e\in\cW_\cB}Z_{s(e)}Z_{t(e)}
\;\longmapsto\;
+\sum_{e\in\cW_\cB}Z_{s(e)}Z_{t(e)}\,.
\end{equation}
After then half-gauging this configuration in $\cB$ we obtain
\begin{equation}
    H_{\eta(\cW_\cB),\cD(\cW)} = -\sum_{e\in\cA}Z_{s(e)}Z_{t(e)} - \sum_{v\in\cA}X_v - \sum_{e\in\cB/\cW_\cB}X_e-\sum_{v\in\cB}A_v + \sum_{e\in\cW_\cB,v\in( e\cap\cW_\cA)}Z_vX_e\,.
\end{equation}
This Hamiltonian is unitarily equivalent to $H_{\cD(\cW)}$ via $\prod_{e\in\cW_\cB}Z_e$, which acts only along the interface. Therefore, the duality interface absorbs the 0-form symmetry defect and we can write the fusion
\begin{equation}
   \cD(\cW) \otimes \eta_0(\Sigma_1^\vee) \cong \cD(\cW)\,,
    \label{eq:etaDFusion}
\end{equation}
where we use $\cong$ to denote equivalence of configurations that can be moved or fused by unitary operators.

Similarly, the duality interface absorbs the zero-dimensional 1-form symmetry defects $\eta_1(\Sigma_0)$ in the gauged region $\cB$. 
Indeed, $H_{\cD(\cW)}$ is invariant under the defect creation operator $\prod_{e\in \gamma}X_e$ for any semi-infinite path $\gamma$ in $\cB$ that starts at $\Sigma_0 \subset e \in \cW_\cB$.
This implies
\begin{equation}
    \eta_{1}(\Sigma_0)\otimes \cD(\cW)\cong \cD(\cW)\,,
    \label{eq:DetaFusion}
\end{equation}
Analogous fusion rules also hold for the conjugate interface $\cD^\dagger$.

\subsection{Movement Operator}
\label{subsec:0formmovement}
As we reviewed above, an interface or defect is topological if it can be moved by local unitaries. For ordinary symmetry defects, this unitary can be inferred directly from the symmetry operator. For the gauging interface $\cD$, the topological nature is more obscure.
We now make the topological nature of the gauging interface explicit by constructing its unitary movement operator. 
\medskip\noindent
Interfaces at different positions should be related by a unitary, mapping the corresponding  constrained Hilbert spaces and interface Hamiltonians into one another.  Therefore, for any two interface loci $\cW$ and $\cW'$ there should exist a unitary $\Lambda^{0,1}_{\cW,\cW'}$ such that
\begin{equation}
    \Lambda^{0,1}_{\cW,\cW'}H_{\cD(\cW)}\left(\Lambda^{0,1}_{\cW,\cW'}\right)^\dagger = H_{\cD(\cW')}\,.
    \label{eq:2dMovementOperatorConstrained}
\end{equation}
We call $\Lambda^{0,1}_{\cW,\cW'}$ the \textit{movement operator},  in analogy with the $(1+1)d$ case~\cite{Seiberg2024}.

\paragraph{Movement operator on unconstrained Hilbert space.}
Interface Hamiltonians live on different constrained Hilbert spaces for different interface locations. Therefore, it is convenient to embed everything into a common unconstrained Hilbert space and keep track of the constraints explicitly.
We take the unconstrained product Hilbert space
\begin{equation}\label{eq:unconstrainedhilbertspace}
    \cH_{\text{unconstr}} = \Big(\bigotimes_{v\in\Gamma}\C_v^2\Big)\otimes\Big(\bigotimes_{e\in \Gamma} \C_e^2\Big)
\end{equation}
and impose the operator identities\footnote{The edges are in the $X$-basis due to our choice of basis in the gauging procedure. We could have equally chosen the $Z$-basis, then minimal coupling would have been $Z_{s(e)}Z_eZ_{t(e)}$.}
\begin{subequations}
    \begin{align}
        X_e &= 1 \quad \text{for all} \quad e\in\cA\,, \label{eq: constraintA}\\
        Z_v &= 1 \quad \text{for all} \quad v\in\cB\,, \label{eq: constraintB}
    \end{align}
\end{subequations}
together with the modified no-twist constraint
\begin{equation}
    V_\gamma(\cW) = \left(\prod_{e\in\gamma\cap\cA}Z_{s(e)}Z_{t(e)}\right)\left(\prod_{e\in\gamma\cap(\cB/\cW_\cB)}X_e\right)\left(\prod_{e\in\gamma\cap\cW_{\cB},v\in e\cap\cW_\cA}Z_vX_e\right) = 1\,,
    \label{eq:2dConstraintsDualityDefect}
\end{equation}
for any closed, contractible loop $\gamma\in\Gamma$. In particular, we have non-trivial constraints outside of the usual twist sector constraints, which is important in the study of movement operator.

The first two constraints~\eqref{eq: constraintA},\eqref{eq: constraintB} trivialize the edge degrees of freedom in $\cA$ and the vertex degrees of freedom in $\cB$, and the last constraint~\eqref{eq:2dConstraintsDualityDefect} is the half-gauged version of the no-twist condition. Note that if $\gamma\subset\cB$, \eqref{eq:2dConstraintsDualityDefect} makes the $\Z_2^{(1)}$ symmetry topological in $\cB$, while for $\gamma\subset\cA$ it enforces the absence of $\Z_2^{(0)}$ twists.
Since the constraints depend on the interface location, the Hilbert spaces before and after moving are different. Defining the projector onto the constrained subspace
\begin{equation}
    P(\cW) = \prod_{e\in\cA}\frac{1}{2}(1+X_e)\prod_{v\in\cB}\frac{1}{2}(1+Z_v)\prod_{\gamma\in\Gamma}\frac{1}{2}(1+V_\gamma)\,,
\end{equation}
we can rewrite \eqref{eq:2dMovementOperatorConstrained} as
\begin{equation}
    \Lambda_{\cW,\cW'}^{0,1} H_{\cD(\cW)}P(\cW)\left(\Lambda_{\cW,\cW'}^{0,1}\right)^\dagger =  H_{\cD(\mathcal{W'})}P(\cW')\,,
    \label{eq:2dMovementOperatorUnconstrained}
\end{equation}
where (by a slight abuse of notation) we now take $\Lambda$ to be the unitary acting on the unconstrained Hilbert space. 

\begin{figure}
    \centering
    \includegraphics[width=.75\textwidth]{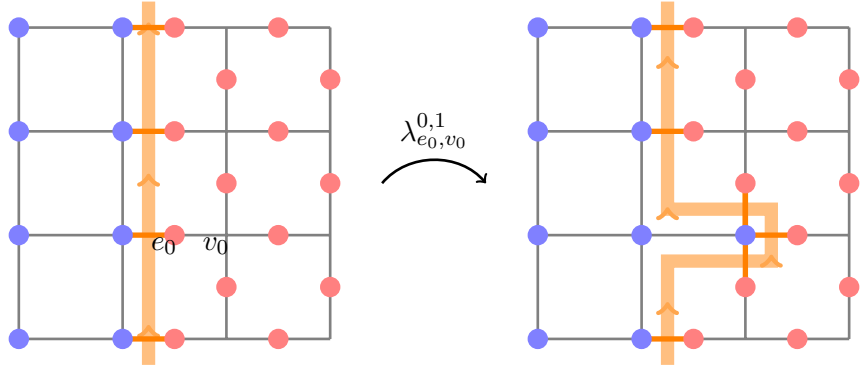}
    \caption{The movement operator $\lambda^{0,1}_{e_0,v_0}$ removes the edge $e_0$ from region $\cB$ and enlarges region $\cA$ by the vertex $v_0$.}
    \label{fig:MovementOperator}
\end{figure}

\paragraph{Construction of movement operator.}
A general movement operator $\Lambda^{0,1}_{\cW,\cW'}$ can be constructed as a product of \textit{minimal} movement operators $\lambda^{0,1}_{e_0,v_0}$. These minimal movement operators move the gauging interface $\cW$ to include an adjacent vertex in the gauged region $\cB$, as illustrated in Figure~\ref{fig:MovementOperator}. On the unconstrained Hilbert space~\eqref{eq:unconstrainedhilbertspace}, they will transform the interface Hamiltonian and the constraints simultaneously. Therefore, the \textit{minimal} movement operator $\lambda^{0,1}_{e_0,v_0}$ implements the local ungauging of a vertex $v_0$, by adding and removing the relevant constraints in the constrained Hilbert space.   

\begin{figure}
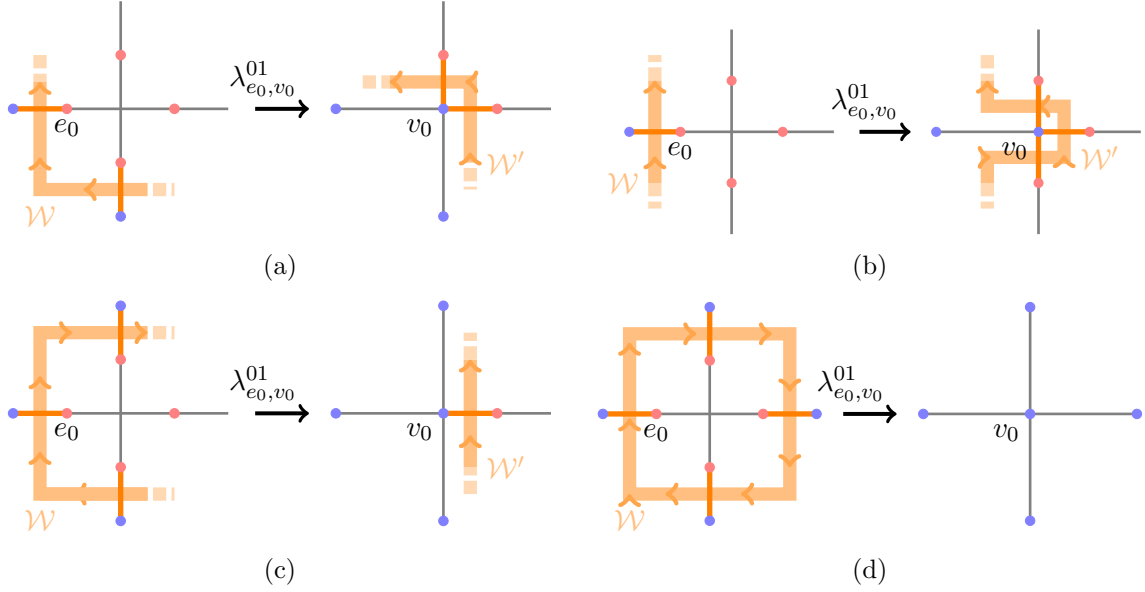

    \centering

    \begin{subfigure}[b]{0.48\textwidth}
        \centering
        \includegraphics[width=\linewidth]{Figures/tikz/Mvmnt1_simple.tikz}
        \caption{} 
        \label{fig:2dMovement1main}
    \end{subfigure}\hfill
    \begin{subfigure}[b]{0.48\textwidth}
        \centering
        \includegraphics[width=\linewidth]{Figures/tikz/Mvmnt2_simple.tikz}
        \caption{} 
        \label{fig:2dMovement2main}
    \end{subfigure}

    \vspace{0.5em} 

    \begin{subfigure}[b]{0.48\textwidth}
        \centering
        \includegraphics[width=\linewidth]{Figures/tikz/Mvmnt3_simple.tikz}
        \caption{} 
        \label{fig:2dMovement3main}
    \end{subfigure}\hfill
    \begin{subfigure}[b]{0.48\textwidth}
        \centering
        \includegraphics[width=\linewidth]{Figures/tikz/Mvmnt4_simple.tikz}
        \caption{} 
        \label{fig:2dMovement4main}
    \end{subfigure}

    \caption{Up to rotations, there are four possible local movements in 2d. These are realized by the unitary $\lambda^{01}_{e_0,v_0}$ which removes an edge $e_0$ and adds the vertex $v_0$. Additionally, there may be constraints that trivialize after the movement, possibly projecting out other edges.}
    \label{fig:2dMovementsmain}
\end{figure}

We propose that in $d=2$, such a minimal movement operator takes the form
\begin{equation}
    \lambda_{e_0,v_0}^{0,1} = \left(\prod_{e\ni v_0,e\neq e_0} \mathsf{CZ}_{e,v_0}\right)\mathsf{H}_{v_0}\mathsf{S}_{e_0,v_0}, \quad e_0\in\cW_{\cB},v_0= e_0\cap \mathcal{\cB}\,,
    \label{eq:MinimalMovement2d}
\end{equation}
where $\mathsf{H}_{v_0}$ is the Hadamard gate acting on $v_0$ and $\mathsf{S}_{e_0,v_0}$ is the swap gate that exchanges the degrees of freedom on $e_0$ and $v_0$.
This is the natural two-dimensional generalization of the $d=1$ movement operator~\cite{Aasen2016,Seiberg2024,Okada2024}\footnote{In the cited works, the authors already include a half-translation in their definition of gauging, and thus define their movement operator without the swap gate.}
\begin{equation}
    \lambda_{e_0,v_0}^{0,0} = \mathsf{CZ}_{v_0,e_0+1}\mathsf{H}_{v_0}\mathsf{S}_{e_0,v_0}\,.
    \label{eq:1dMovement}
\end{equation}
As depicted in Figure~\ref{fig:2dMovementsmain}, in $(2+1)d$, locally around a vertex a generic interface can take four different shapes (up to rotations). It is enough to check our proposal~\eqref{eq:MinimalMovement2d} by verifying these four types of movement, and the details are found in Appendix~\ref{app:Movement}.

For the alternative interface presentation~\eqref{eq:0formAlternativeInterface}, the corresponding minimal movement operator is
\begin{equation}
    \widetilde{\lambda}^{0,1}_{e_0,v_0} 
    = U_{\cW'}\,\lambda_{e_0,v_0}^{0,1}\,U_{\cW}^\dagger\,,
    \label{eq:AlternativeMovement}
\end{equation}
where $U_\cW$ is defined in~\eqref{eq:2dRelationBtwPresentations} and $\cW'$ is the interface after the move.

\section{Condensation Defects and Fusions in $(2+1)d$}
\label{sec:0formCond}
Although the duality interfaces $\mathcal{D}$ and $\mathcal{D}^\dagger$ interpolate between two different theories, they still serve as building blocks for more general non-invertible symmetry defects living within a given theory. 
One important example is \emph{condensation defects}, which have been studied extensively in continuum field theory~\cite{Roumpedakis2023,Choi2023,Cordova:2024mqg}.
Here, we study condensation defects on the lattice by gauging a connected finite sublattice $\Sigma$ of the spatial lattice. We give explicit lattice defect Hamiltonians featuring such defects and verify their fusion rules, in agreement with the continuum analysis.

More concretely, condensation defects are obtained by inserting duality interfaces along $\partial\Sigma$. Therefore,
\begin{equation}
    \cC_0(\Sigma) \;\cong\; \cD(\partial\Sigma)\,, \qquad  \cC_{1}(\Sigma) \;\cong\; \cD^\dagger(\partial\Sigma)\,,
\end{equation}
where $\cC_{p}$ denote the condensation defect for $\mathbb Z_2^{(p)}$ form symmetry.~\footnote{Here we abuse the terminology a bit because $\cC_{0}$ is the non-simple projection defect.}
We call two condensation defects equivalent if they can be continuously deformed into each other using the movement operator.
In $(2+1)d$, there are two non-equivalent cases: $\Sigma$ is fully contractible, or it carries one non-contractible cycle.

\subsection{Condensation Defects of $\Z_2^{(0)}$ Symmetry.}
In the TFIM~\eqref{eq:TFIM}, gauging $\Z_2^{(0)}$ on $\Sigma$ yields the defect Hamiltonian with a condensation defect $\cC_0(\Sigma)$
\begin{equation}
    H_{\cC_0(\Sigma)} 
    = -\sum_{e\in\Gamma/\Sigma}Z_{s(e)}Z_{t(e)}
      -\sum_{v\in\Gamma/\Sigma}X_v
      -\sum_{v\in\Sigma}A_v
      -\sum_{e\in\Sigma/\partial\Sigma}X_e
      -\sum_{e\in\partial\Sigma}\Big(\prod_{v\in e\cap(\Gamma/\Sigma)}Z_v\Big)X_e\,.
\end{equation}

If $\Sigma$ is fully contractible, through movement operators, this defect Hamiltonian is equivalent to the one with the defect localized on a single edge $\Sigma_0$
\begin{equation}
    \Lambda_{\Sigma,\Sigma_0} H_{\cC_0(\Sigma)} \Lambda^\dagger_{\Sigma,\Sigma_0}  = H_{\cC(\Sigma_0)} = -\sum_{e\in\Gamma/\Sigma_0}Z_{s(e)}Z_{t(e)}-\sum_{v\in\Gamma}X_v- Z_{s(\Sigma_0)}X_{\Sigma_0}Z_{t(\Sigma_0)}\, ,
\end{equation}
together with the constraint $X_{\Sigma_0}=1$ fixing this extra qubit. Again, this constraint comes from tracking how the no-twist constraints behave after gauging. Therefore,
\begin{equation}
    \cC_{0}(\Sigma)\cong\cC_0(\Sigma_0)\cong 1,\quad \text{for contractible }\Sigma.
\end{equation}

If $\Sigma$ carries one non-contractible cycle, through movement operators, this defect Hamiltonian is equivalent to one where the defect is localized on a one-dimensional cycle $\Sigma_1^\vee$ in the dual lattice:
\begin{equation}
    \Lambda_{\Sigma,\Sigma_1^\vee} H_{\cC_0(\Sigma)} \Lambda^\dagger_{\Sigma,\Sigma_1^\vee}  = H_{\mathcal{C}_0(\Sigma_1^\vee)} = - \sum_{e\notin\Sigma_1^\vee}Z_{s(e)}Z_{t(e)} -\sum_v X_v - \sum_{e\in\Sigma_1\vee}Z_{s(e)}X_eZ_{t(e)}\,,
    \label{eq:ConstraintCondensationDefect}
\end{equation}
together with additional constraints $V_\gamma$ (see Eq.~\eqref{eq:2dConstraintsDualityDefect})
\begin{equation}
    \tikzCondConstraint =1\,,
\end{equation}
for any two edges in $\Sigma_1$. 
This fixes the extra edge degrees of freedom along $\Sigma_1^\vee$ down to one, say $X_{\Sigma_1^\vee}$.
Physically, the two eigenvalues of this ancilla qubit $X_{\Sigma_1}=\pm1$ label different defect channels: whether or not there is a $\Z_2^{(0)}$ defect $\eta_0(\Sigma_1^\vee)$ inserted along $\Sigma_1^\vee$. The defect Hamiltonian can be diagonalized as
\begin{equation}
    H_{\mathcal{C}_0(\Sigma_{1}^\vee)} = H\ket{0}_{\Sigma_{1}^\vee}\bra{0} \oplus H_{\eta_0(\Sigma_{1})}\ket{1}_{\Sigma_{1}^\vee}\bra{1}\,,
\end{equation}
where $H_{\eta_0(\Sigma_1^\vee)}$ is the TFIM Hamiltonian with a $\mathbb{Z}_2^{(0)}$ defect $\eta_0$ inserted along $\Sigma_1^\vee$. Therefore,
\begin{equation}
    \cC_0(\Sigma) \cong \cC_0(\Sigma_1) \cong 1\oplus\eta_0(\Sigma_1),\quad \text{for non-contractible }\Sigma\,.
\end{equation}
Because $\partial\Sigma_1 = \partial\Sigma_1^L\sqcup\partial\Sigma_1^R$ has two components of opposite orientation, we also have
\begin{equation}
    \cD^\dagger(\partial\Sigma_1^R) \otimes \cD(\partial\Sigma_1^L) \cong \cC_0(\Sigma_1) \cong  1\oplus\eta_0(\Sigma_1)\,,
    \label{eq:decommpC0}
\end{equation}
which is the defect analogue of the operator fusion rule~\eqref{eq:fusionop2d}.

\subsection{Condensation Defects of $\Z_2^{(1)}$ Symmetry.}
In the $\Z_2$ lattice gauge theory~\eqref{eq:2dLGT}, gauging $\Sigma$ in the dual theory yields a defect Hamiltonian with a condensation defect $\cC_{1}(\Sigma)$ 
\begin{equation}
H_{\cC_{1}(\Sigma)} 
= -\sum_{v\in \Gamma/\Sigma} A_v 
  - \sum_{e\in \Gamma/\Sigma} X_e 
  -\sum_{e\in\Sigma}Z_{s(e)}Z_{t(e)} 
  - \sum_{v\in\Sigma/\partial\Sigma} X_v 
  - \sum_{v\in\partial\Sigma}\Big(\prod_{e\ni v,\, e\in \Gamma/\Sigma}Z_e\Big)X_v.
\end{equation}
For a contractible sublattice, the simplest nontrivial example is obtained by gauging a single vertex $\Sigma_0=v_0$:
\begin{equation}
    H_{\mathcal{C}_{1}(\Sigma_0)} = -\sum_{v\neq \Sigma_0}A_v - \sum_e X_e -A_{\Sigma_0}X_{\Sigma_0} = H^{(1)}\ket{0}\bra{0} \oplus H^{(1)}_{\eta_{1}(\Sigma_0)}\ket{1}\bra{1}\,.
    \label{eq:HC1decomp}
\end{equation}
Hence
\begin{equation}
    \mathcal{C}_{1}(\Sigma_0) = 1\oplus\eta_{1}(\Sigma_0)\,.
\end{equation}
Using movement operators, gauging any contractible sublattice $\Sigma$ can be unitarily deformed into the above Hamiltonian.

More interesting is the case where the gauged sublattice carries one non-contractible cycle, say $\Sigma_1$. 
The corresponding condensation defect is now one-dimensional:
\begin{align}
    H_{\cC_{1}(\Sigma_1)} &= - \sum_{v\notin\Sigma_1}A_v-\sum_{e\notin \Sigma_1}X_e - \sum_{e\in\Sigma_1}Z_{s(e)}Z_{t(e)} - \sum_{v\in\Sigma_1} X_v\left(\prod_{e\ni v,e\notin \Sigma_1}Z_e\right)\\
    &= - \sum_{v\notin\Sigma_1}\tikzAv-\sum_{e\notin \Sigma_1}\tikzXe - \sum_{e\in\Sigma_1}\tikzZZonC- \sum_{v\in\Sigma_1} \tikzZXZonC.
    \label{eq:HC1d2}
\end{align}
This defines an indecomposable one-dimensional defect $\cC_1(\Sigma_1)$~\footnote{There is no way to decompose $\cC_1(\Sigma_1)$ into a sum of simple symmetry defects because the only other simple defects in the $\Z_2^{(1)}$ theory are the point-like defects $\eta_1$ .}, built as a fine mesh of point-like $\eta_1$ defects, with an emergent one-dimensional $\Z_2^{(0)}$ symmetry $\prod_{v\in\Sigma_1}X_v$ along $\Sigma_1$. 
On the defect, there lives a $(1+1)d$ TFIM with transverse field $X$ coupled to the bulk. This means that the $\mathbb{Z}_2^{(1)}$ Wilson lines can end on the condensation defect: Inserting a Wilson line along a 1d line $\gamma$ amounts to modifying $Z_e\to -Z_e$ for $e\in \gamma$. This creates excitations at the ends of $\gamma$. However, now we see that if $\gamma$ ends on the condensation defect, we can cancel this excitation by a point-like spin-flip of the $(1+1)d$ TFIM. Therefore, we conclude that bulk Wilson lines can end on the condensation defect where they become the spin-flip operator $Z_v$.

\subsection{Defect Fusion Rules}
Given that condensation defects are created from gauging, they must be able to absorb the corresponding symmetry defects like~\eqref{eq:etaDFusion} and \eqref{eq:DetaFusion}. From the decomposition of $\mathcal{C}_0(\Sigma_1)$ \eqref{eq:decommpC0}, it follows directly that 
\begin{align}
    \begin{split}
    \eta_0(\Sigma_{1})\otimes  \cC_0(\Sigma_{1}) &\cong \cC_0(\Sigma_{1})\,.
    \label{eq:FusionEtaC2d1}
    \end{split}
\end{align}
As explained above, for $\mathcal{C}_1(\Sigma_1)$, moving an $\eta_1(\Sigma_0)$-defect onto this condensation defect at site $v_0$ results in the local change of sign
\begin{equation}
    - X_{v_0}\prod_{e\ni v_0,e\in\Sigma_1} Z_e \longrightarrow +X_{v_0}\prod_{e\ni v_0,e\in\Sigma_1} Z_e\,,
\end{equation}
which can be canceled by applying the unitary $Z_{v_0}$. Note that this changes the symmetry eigenvalue of the $\Z_2^{(0)}$-symmetry (generated by $\prod_{v\in\Sigma_1}X_v$)  on the condensation defect. This shows that the non-trivial $\mathcal{C}_1(\Sigma_1)$ condensation defect can indeed absorb the bulk $\Z_2^{(1)}$-defects, 
\begin{align}
    \begin{split}
    \eta_{1}(\Sigma_0) \otimes \mathcal{C}_{1}(\Sigma_1) &\cong \mathcal{C}_{1}(\Sigma_1)\,,
    \label{eq:FusionEtaC2d1}
    \end{split}
\end{align}
at the expense of creating a charge on the condensation defect.

\paragraph{Fusion with condensation defects.}
The self-fusion of the decomposable condensation defects is straightforward:
\begin{equation}
\mathcal{C}_0(\Sigma_{1})\otimes\mathcal{C}_0(\Sigma_{1}) \cong (1\oplus\eta_0(\Sigma_{1}))\otimes(1\oplus\eta_0(\Sigma_{1}))\cong2\mathcal{C}_0(\Sigma_{1})\,.
\end{equation}
and similarly
\begin{equation}
   \mathcal{C}_1(\Sigma_0)\otimes \cC_1(\Sigma_0) \cong 2\cC_1(\Sigma_0)\,.
\end{equation}

\begin{figure}
    \centering
    \includegraphics[width=0.8\linewidth]{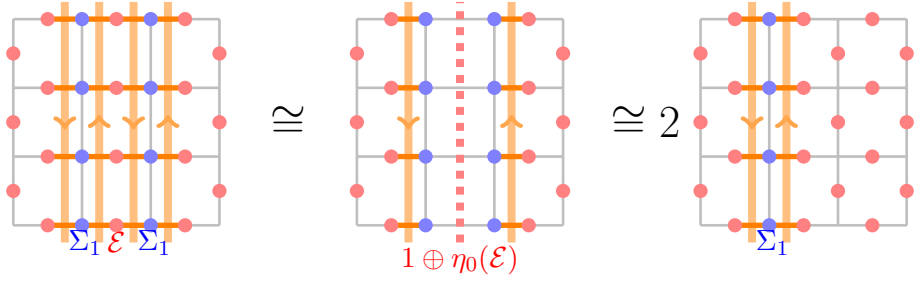}
    \caption{Fusing two $\cC_1$-defects located on $\Sigma_1$ and $\Sigma_1'$.}
    \label{fig:C1C1Fusion}
\end{figure}

For the indecomposable one-dimensional condensation defect $\cC_{1}(\Sigma_1)$ we can express its self-fusion using the duality interfaces, as illustrated in Figure \ref{fig:C1C1Fusion}. The Hamiltonian with two $\cC_{1}$ defects on $\Sigma_1$ and $\Sigma_1'$ reads
\begin{equation}
    H_{C_1(\Sigma_1),C_1(\Sigma_1')} = -\sum_{v\in\mathcal{A},\mathcal{B}}A_v-\sum_{e\in\mathcal{A},\mathcal{B}}X_e - \sum_{e\in\Sigma_1,\Sigma_1'}Z_{s(e)}Z_{t(e)} - \sum_{v\in\Sigma_1,\Sigma_1'}ZX_vZ-\sum_{e\in\mathcal{E}}X_e\,,
\end{equation}
where $\mathcal{A}$ ($\mathcal{B}$) are the regions to the left (right) of the condensation defects, and $\mathcal{E}$ are the edges in between.  Applying the unitary $U=\prod_{e\in\mathcal{E}}\prod_{v\in\Sigma_1,\Sigma_1'}\mathsf{CZ}_{v,e}$, we have
\begin{equation}
    H_{C_1(\Sigma_1),C_1(\Sigma_1')} \cong -\sum_{v\in\mathcal{A},\mathcal{B}}A_v-\sum_{e\in\mathcal{A},\mathcal{B}}X_e - \sum_{e\in\Sigma_1,\Sigma_1'}Z_{s(e)}Z_{t(e)} - \sum_{v\in\Sigma_1,\Sigma_1'}ZX_v-\sum_{e\in\mathcal{E}}ZX_eZ\,,
\end{equation}
which is a $C_0$ condensation defect (the last term) sandwiched by two gauging interfaces (the second last term). This is equivalent to 
\begin{equation}
    \mathcal{C}_1(\Sigma_1)\otimes\mathcal{C}_1(\Sigma_1) 
    \;\cong\; 
    \cD^\dagger \otimes (\cD\otimes\cD^\dagger)\otimes\cD 
    \;\cong\;
    \cD^\dagger\otimes \cC_0 \otimes \cD\,.
\end{equation}
where we use $\cC_1(\Sigma_1)\cong \cD^\dagger(\partial\Sigma_1^L)\otimes\cD(\partial\Sigma_1^R)$ and associativity of fusion.
Since $\cC_0\cong 1\oplus\eta_0$ and $\eta_0$ fuses trivially through $\cD$ according to~\eqref{eq:etaDFusion}, we obtain
\begin{equation}
    \mathcal{C}_1(\Sigma_1)\otimes\mathcal{C}_1(\Sigma_1) \cong 2\,\mathcal{C}_1(\Sigma_1)\,.
\end{equation}

We can also derive the fusion rule between the condensation defect and the duality interface $\mathcal{D}$. 
Consider the usual setup of a split lattice $\Gamma = \mathcal{A}\sqcup \mathcal{B}$ with a $\Z_2^{(0)}$-symmetry in $\mathcal{A}$, a $\Z_2^{(1)}$-symmetry in $\mathcal{B}$, and a duality interface $\mathcal{D}(\mathcal{W})$ in between. Inserting a condensation defect $\mathcal{C}_0(\Sigma_1)$ with $\Sigma_1\subset \mathcal{A}$ leads to 
\begin{equation}
    \mathcal{C}_0(\Sigma_1) \otimes \mathcal{D}(\mathcal{W}) \cong 2\mathcal{D}(\mathcal{W})
\end{equation}
because of the fusion~\eqref{eq:etaDFusion}. If we insert the indecomposable $\mathcal{C}_1(\Sigma_1)$ for $\Sigma_1\subset \mathcal{B}$, we have
\begin{equation}\label{eq:condensationinterfacecommut}
    \cD(\cW)\otimes\cC_1(\Sigma_1)  \cong 2\mathcal{D}(\mathcal{W})
\end{equation}
because the following Hamiltonians
\begin{align}
    \begin{split}
        H_{\cD(\cW),\cC_1(\Sigma_1)} &=-\sum_{e\in\cA}Z_{s(e)}Z_{t(e)} - \sum_{v\in\cA}X_v- \sum_{v\in\cB} A_v - \sum_{e\in\cB/\cW_{\cB}}X_e  \\&\quad -\sum_{e\in \cW_{\cB}, v\in e\cap\cW_{\cA}}Z_vX_e
         -\sum_{v\in\Sigma_1} X_v\prod_{e\ni v,e\notin\Sigma_1}Z_e - \sum_{e\in\Sigma_1}Z_{s(e)}Z_{t(e)}
    \end{split}
\end{align}
and
\begin{align}
    \begin{split}
        H_{\cC_0(\Sigma_1'),\cD(\cW')} &= -\sum_{e\in\cA'/\Sigma_1'}Z_{s(e)}Z_{t(e)} - \sum_{v\in\cA'}X_v- \sum_{v\in\cB'} A_v - \sum_{e\in\cB'/\cW_{\cB}'}X_e\\
        &\quad  -\sum_{e\in \cW'_{\cB}, v\in e\cap\cW_{\cA}'}Z_vX_e - \sum_{e\in\Sigma_1'}X_e\prod_{v\in e}Z_v
    \end{split}
\end{align}
are related by the unitary $\prod_{e\in\cW_\cB}\prod_{v\ni e}\mathsf{CZ}_{v,e}$ (and possible unitary movements, depending on the location).

\subsection{Relating the Different Condensation Defects.}
\label{Subsec:RelatingCondDefects}
\begin{figure}
    \centering
    \includegraphics[width=\linewidth]{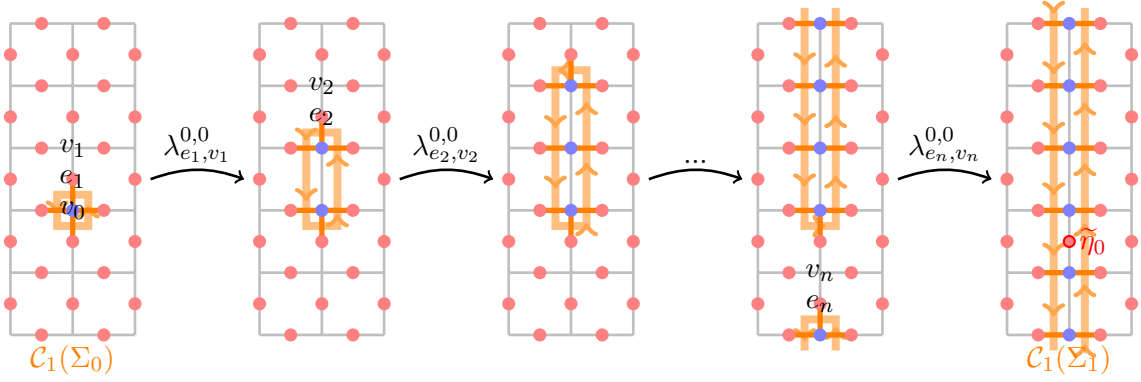}
    \caption{After successive application of the movement operator, we can relate the decomposable defect $\mathcal C_1(\Sigma_0)$ to the indecomposable $\mathcal C_1(\Sigma_1)$.}
    \label{fig:1dMovementCondensationDefect}
\end{figure}

With the movement operators, we can progressively enlarge the single vertex $\Sigma_0$ to the non-contractible cycle $\Sigma_1$, thus relating the decomposable defect $\cC_1(\Sigma_0)$ to the indecomposable defect $\cC_1(\Sigma_1)$. The whole process is illustrated in Figure~\ref{fig:1dMovementCondensationDefect}. This construction can also be seen as the first step in obtaining the gauging map from moving the interface, which will be done in Subsection \ref{subsec:2dOperatorFromDefect}.

\medskip\noindent Recall the Hamiltonian of a single $\cC_1(\Sigma_0)$- defect
\begin{equation}
    H_{\cC_1(\Sigma_0)} = -\sum_{v\neq \Sigma_0}A_v - \sum_e X_e -A_{\Sigma_0}X_{\Sigma_0}
\end{equation}
We label the vertices and edges on $\Sigma_1$ by $\Sigma_1 = \{e_0,v_0,e_1,v_1,...,e_n,v_n\}$.  Then, we first apply the unitary $\mathsf{CZ}_{v_0,e_0}\mathsf{CZ}_{v_0,e_1}$ and then we successively apply the incremental one-dimensional movement operator $\lambda_{e,v}^{0,0} = \mathsf{CZ}_{v,e+1}\mathsf{H}_{v}\mathsf{S}_{e,v}$, as defined in Eq.~\eqref{eq:1dMovement}. This results in the Hamiltonian
\begin{align}
    \begin{split}
    H_{\cC_1(\Sigma_0)} &\cong -\sum_{v\notin\Sigma_1}A_v-\sum_{e\notin\Sigma_1}X_e-\sum_{e\in(\Sigma_1/e_0)}Z_{s(e)}Z_{t(e)}-\sum_{v\in\Sigma_1}X_v\left(\prod_{e\ni v,e\notin \Sigma_1}Z_e\right)- Z_{v_0}X_{e_0}Z_{v_n}\,,\\ &= H_{\cC_1(\Sigma_1)}\oplus H_{\cC_1(\Sigma_1),\widetilde{\eta}_{0}(e_0)}\,,
    \end{split}
\end{align}
which is made up of a one-dimensional condensation defect $C_1(\Sigma_1)$, together with an additional single additional degree of freedom $X_{e_0}$. This corresponds to a symmetry defect of the $\Z_2^{(0)}$-symmetry on the condensation defect $\cC_1(\Sigma_1)$. Written in terms of defects, we get
\begin{equation}
    1\oplus \eta_1(\Sigma_0) \cong \cC_1(\Sigma_0) \cong \cC_1(\Sigma_1)(1\oplus\widetilde{\eta}_0(\Sigma^\vee_0))\,.
\end{equation}
where we denoted $e_0=\Sigma_0^\vee$. After projection onto the trivial defect channel, we get the indecomposable condensation defect $\cC_1(\Sigma_1)$. We can package the above discussion into the operator
\begin{equation}
    \mathsf{C}_1^{1\oplus \widetilde\eta_0}(\Sigma_1) = \left( \prod_{i\in{1,...,n}}^\leftarrow\lambda^{0,0}_{e_{i},v_i}\right) \mathsf{CZ}_{v_0,e_0}\mathsf{CZ}_{v_0,e_1}\ket{+}_{v_0}\,,
    \label{eq:2dCondOperator}
\end{equation}
where $\overset{\leftarrow}{\prod}$ means that we write the product from right to left (such that $\lambda_{e_1,v_1}^{00}$ is applied first). Then, we can also define the creation operator of an indecomposable condensation defect as
\begin{equation}
    \mathsf{C}_1(\Sigma_1) = \bra{+}_{e_0}\mathsf{C}_1^{1\oplus \widetilde\eta_0}(\Sigma_1)\,.
\end{equation}
Notice that this operator coincides with the 1d $\Z_2^{(0)}$ gauging operator derived in \cite{Seiberg2024,Okada2024}. Therefore, this confirms the common lore that condensing a 1-form symmetry in 2d is analogous to gauging a 0-form symmetry in 1d.

\subsection{Gauging Map From Moving Interfaces}
\label{subsec:2dOperatorFromDefect}
In $(1+1)d$, it was shown in~\cite{Seiberg2024,Okada2024} that the Kramers--Wannier gauging operator can be reconstructed by starting from a decomposable condensation defect, made up of two duality interfaces, then moving one of the duality interfaces once around the entire spatial circle using the movement operator, and finally projecting out the condensation defect again.
This gives an alternative and very efficient realization of the gauging map in terms of a single environmental degree of freedom, as opposed to the usual gauging map, made up of an extensively large environment, as in \eqref{eq:2dGaugingMap}.
\begin{figure}
    \centering
    \includegraphics[width=1.1\linewidth]{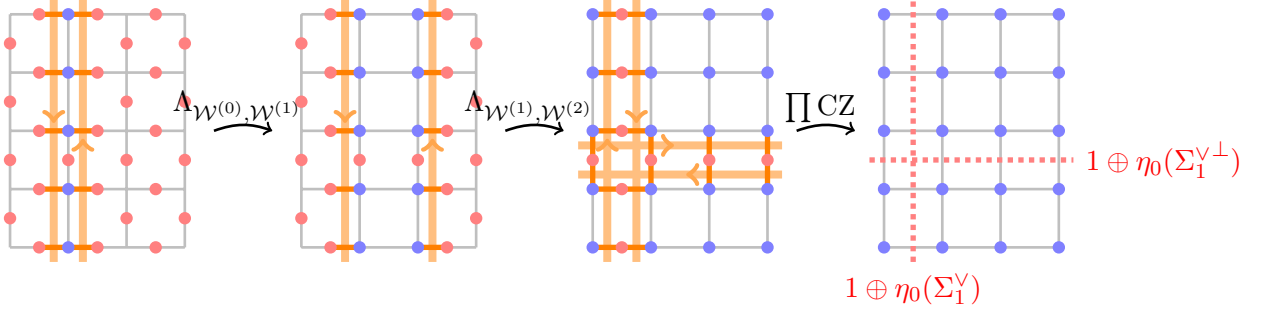}
    \caption{Starting from a one-dimensional condensation defect $\cC_1(\Sigma_1)$ in the $\Z_2^{(1)}$-theory (and its twisted sector), we use the unitary movement operator to progressively enlarge it until the gauged region fills the entire lattice. Here, we defined $\cW ^{(i)}$ to be the right boundary of the condensation defect after the $i$-th iteration of the movement.}
    \label{fig:DualityFromMovement}
\end{figure}

\medskip\noindent
Here, we show that the same logic extends to $(2+1)d$. We begin in the $\Z_2^{(1)}$ symmetric theory with a single extra degree of freedom at $v_0=\Sigma_0$, representing a decomposable $0$-dimensional condensation defect $\cC_1(\Sigma_0)=1\oplus\eta_1(\Sigma_0)$.
Using the movement operator, we first extend this to a one-dimensional condensation defect $\cC_1(\Sigma_1)$, as shown in Fig.~\ref{fig:1dMovementCondensationDefect}. Then we further sweep the gauging interfaces around the torus, and fuse the parallel gauging interfaces, as sketched in Fig.~\ref{fig:DualityFromMovement}. We show that this produces two 0-form defects, around the two non-contractible cycles of the torus.
Projecting onto the untwisted sector yields a duality operator $\mathsf{D}^\dagger$, implementing the gauging of $\mathbb Z_2^{(1)}$ symmetry. The details of this calculation are presented in Appendix~\ref{app:Movement}.
Thus, also in $(2+1)d$, the gauging map can be viewed as generated by moving a gauging interface around the spatial manifold in the presence of an appropriate condensation defect. This generalizes the notion that non-invertible symmetries act by quantum operations, that was put forward in the $(1+1)d$ case \cite{Okada2024}. The operator takes the explicit form
\begin{equation}
    \mathsf{D}^{\dagger} = \bra{+}_{\Sigma_1^\vee}\bra{+}_{{\Sigma_1^\vee}^\perp}\left(\prod_{i=0,...,L-1}^{\leftarrow}\Lambda^{0,1}_{\cW^{(i)},\cW^{(i+1)}}\right)\left(\prod_{v\in \Sigma_1}\prod_{e\ni v,e\notin \Sigma_1}\mathsf{CZ}_{v,e}\right)\mathsf{C}_1(\Sigma_1)^{1\oplus \widetilde\eta_0}
\end{equation}
where $\mathsf{C}_1$ is the condensation defect creation operator, as defined in Eq.~\eqref{eq:2dCondOperator} and
\begin{equation}
    \Lambda^{0,1}_{\cW^{(i)},\cW^{(i+1)}} = \prod_{e\in\cW^{(i)}}\lambda_{e,t(e)}^{0,1} \,.
\end{equation}
moves the right interface one lattice site to the right.
In Appendix~\ref{app:Movement}, we work out explicitly that this operator indeed realizes the correct gauging.

\section{Interfaces and Defects for $\Z_2^{(1)}$-Symmetry in $(3+1)d$}
\label{sec:1form}
In this section, we construct lattice gauging maps and the associated topological interfaces that arise from gauging a global $\Z_2^{(1)}$ symmetry in $(3+1)d$.
As the $(3+1)d$ analogue of the Kramers–Wannier duality in $(1+1)d$, gauging the 1-form symmetry defines the Wegner duality of the $\Z_2$ lattice gauge theory.
At the self-dual point, this Wegner duality becomes a non-invertible symmetry,  as recently explored in~\cite{Gorantla2024,Moradi2025,Gorantla2024_2}.
We give the first explicit movement-operator construction of the lattice Wegner duality interface, which becomes a \textit{symmetry defect} at the self dual point, and the associated condensation defects for $\mathbb Z_2^{(1)}$ in $(3+1)d$.
\subsection{Gauging on the Lattice}
We begin by reviewing the gauging of $\mathbb Z_2^{(1)}$ symmetry and the corresponding gauging map on a cubic lattice $\Gamma$ in three spatial dimensions.
The discussion closely parallels the $(2+1)d$ $\mathbb Z_2^{(0)}$ symmetry case in Sec.~\ref{sec:0form}, with vertices and edges exchanged for edges and plaquettes.

\paragraph{$\Z_2^{(1)}$ symmetric lattice models.} 
To each edge $e$ of $\Gamma$ we associate a qubit Hilbert space $\C^2_e$, so that
\begin{equation}
    \cH = \bigotimes_{e\in\Gamma} \C^2_e,
\end{equation}
with Pauli operators $X_e$ and $Z_e$ acting on edge $e$.
We impose the ``zero-flux'' constraint
\begin{equation}\label{eq:zeroflux3d}
    W_{\partial \hat{\Sigma}_3} = \prod_{e\in \partial \hat{\Sigma}_3} X_e = 1,
    \qquad \forall\, \partial \hat{\Sigma}_3 \subset \hat{\Gamma},
\end{equation}
where $\hat{\Gamma}$ is the dual cubic lattice and $\partial \hat{\Sigma}_3$ denotes contractible closed surfaces on $\hat{\Gamma}$.
On this constrained Hilbert space, the topological $\Z_2^{(1)}$ symmetry is generated by Wilson surface operators
\begin{equation}
    W_{\hat{\Sigma}_2} = \prod_{e\in \hat{\Sigma}_2} X_e,
\end{equation}
for any closed surface $\hat{\Sigma}_2$ on the dual lattice. The above constraint~\eqref{eq:zeroflux3d} ensures that $W_{\hat{\Sigma}_2}$ depends only on the homology class of $\hat{\Sigma}_2$.

Local $\Z_2^{(1)}$ invariant operators are generated by
\begin{equation}
    X_e, \qquad B_p = \prod_{e\in p} Z_e,
\end{equation}
where $p$ runs over plaquettes of $\Gamma$.
The algebra generated by these operators (together with the constraints) describes the space of $\Z_2^{(1)}$ symmetric Hamiltonians.
A canonical example is the pure $\Z_2$ lattice gauge theory
\begin{equation}
    H^{(1)}(g) = -\sum_{p\in\Gamma} B_p - g\sum_{e\in\Gamma} X_e.
    \label{eq:Z2LGT}
\end{equation}
As in the $(2+1)d$ case, it is convenient to introduce operators that measure twisted sectors.
For any closed surface $\Sigma_2$ on the direct lattice, we define
\begin{equation}
    V_{\Sigma_2} = \prod_{p\in \Sigma_2} B_p,
\end{equation}
which measures the $\Z_2^{(1)}$ twist through $\Sigma_2$.
In the untwisted sector we set $V_{\Sigma_2} = 1$ for all contractible surfaces $\Sigma_2$, and more generally, the eigenvalues of $V_{\Sigma_2}$ label different twist sectors.
These operators will serve as book-keeping devices that track how the twisted sectors are mapped under gauging.

\begin{figure}
    \centering
    \includegraphics[width=0.6\linewidth]{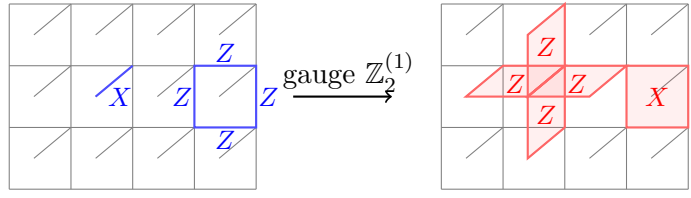}
    \caption{Local action of the gauging map on the three-dimensional lattice.}
    \label{fig:3dlattice}
\end{figure}

\paragraph{Gauging the $\Z_2^{(1)}$ symmetry.} To gauge the $\Z_2^{(1)}$ symmetry, we introduce $\Z_2$ gauge degrees of freedom on the plaquettes $p$ and consider the extended Hilbert space
\begin{equation}
    \cH_{\text{ext}} = \bigotimes_{e\in\Gamma} \C^2_e \bigotimes_{p\in\Gamma} \C^2_p,
\end{equation}
with Pauli operators $X_p, Z_p$ acting on the plaquette qubits.
Minimal coupling promotes the global $\Z_2^{(1)}$-invariant operators to gauge-invariant ones according to
\begin{equation}
    \{X_e,\, B_p,\, W_{\hat{\Sigma}_2},\, V_{\Sigma_2}\}
    \longmapsto
    \{X_e,\, B_p X_p,\, W_{\hat{\Sigma}_2},\, V_{\Sigma_2} \prod_{p\in\Sigma_2} X_p\}\,,
\end{equation}
Gauge transformations are generated by local Gauß operators
\begin{equation}
    G_e = X_e A_e, 
    \qquad A_e = \prod_{p\ni e} Z_p, 
    \qquad \forall e\in\Gamma,
\end{equation}
and physical states obey $G_e = 1$ for all $e$. We apply the unitary
\begin{equation}
    \mathcal{M} 
    = \prod_e \prod_{p\supset e} \mathsf{CZ}_{e,p}:\quad
    X_e \mapsto X_e A_e,\quad
    Z_e \mapsto Z_e,\quad
    X_p \mapsto X_p B_p,\quad
    Z_p \mapsto Z_p\,,
    \label{eq:GaussUnitaries3d}
\end{equation}
to simplify the Gauß law $\mathcal{M} G_e \mathcal{M}^\dagger = X_e=1$, which freezes the edge degrees of freedom and projects onto the gauge invariant sector. Finally we obtain an effective theory purely on the plaquettes.

On this physical subspace, the original symmetry and twist generators are mapped to the gauged ones:
\begin{equation}
    \{X_e,\, B_p,\, W_{\hat{\Sigma}_2},\, V_{\Sigma_2}\}
    \longmapsto
    \{A_e,\, X_p,\, V_{\hat{\Sigma}_2},\, W_{\Sigma_2}\},
    \label{eq:1formGaugingMap}
\end{equation}
where $V_{\hat{\Sigma}_2}$ now measures twists of the dual 1-form symmetry in the plaquette variables and
\begin{equation}
    W_{\Sigma_2} = \prod_{p\in\Sigma_2} X_p,
\end{equation}
are Wilson surfaces of the plaquette variables and generate the quantum $\Z_2^{(1)}$ symmetry. This is the direct 1-form analogue of the $(2+1)d$ gauging map~\eqref{eq:gauging map}. Under gauging, the twist operators $V_{\Sigma_2}$ are mapped to the symmetry generators $W_{\Sigma_2}$. 
Applying this gauging procedure to the Hamiltonian~\eqref{eq:Z2LGT} yields
\begin{equation}\label{eq:3+1ddual}
    H^{(1)}_{\text{dual}}(g) 
    = - g\sum_{e}A_e - \sum_{p}X_p\,.
\end{equation}
The gauged Hamiltonian takes the same form as the original one, up to a half-lattice translation $T_{(\frac{1}{2},\frac{1}{2},\frac{1}{2})}$ that identifies plaquettes with dual edges. Thus, gauging the $\Z_2^{(1)}$ symmetry, also known as Wegner duality, sends $g\mapsto 1/g$ in~\eqref{eq:Z2LGT}. At $g=1$ this duality becomes a non-invertible symmetry of the theory. Without loss of generality, we keep $g=1$ in the following discussion.

\paragraph{Gauging map and Wegner duality operator.} 

To realize the Wegner duality operator, we package the above gauging procedure into a linear map~\cite{Gorantla2024} within the same Hilbert space
\begin{align}
     \mathsf{D}= \otimes_e \bra{+}_e \,
  \mathcal{M}\, \otimes_p \ket{+}_{p}, \label{Wegner op in 3+1}
\end{align}
where we first couple to the plaquette gauge field through adding the product state $\otimes_p \ket{+}_{p}$, then apply the unitary $\mathcal{M}$ and project onto the gauge invariant sector with $\otimes_e \bra{+}_e$. One may check straightforwardly that this map realizes the following transformations
\begin{subequations}
    \begin{align}
        \mathsf{D}X_{e} &= A_e \mathsf{D}, \\
        \mathsf{D}B_{p} &= X_p\mathsf{D},
    \end{align}
\end{subequations}
depicted in Figure \ref{fig:3dlattice}. The fusion of such an operator \cite{Gorantla2024} can be written in terms of non-contractible surfaces $W_{xy},W_{yz},W_{zx}$ and contractible ones generated by $A_v$
\begin{equation}
  \mathsf{D}^2=\frac{1}{2}(1+W_{xy})(1+W_{yz})(1+W_{zx})\prod_{v} \left(\frac{1+A_v}{2}\right),
 \end{equation}
where $A_v =\prod_{e\ni v} X_e$.

\subsection{Duality Interfaces and Defects from Half-Gauging}
We now construct the Hamiltonian with a Wegner duality interface $\mathcal{D}(\cW)$ by gauging in half of space, following the procedure outlined in Section ~\ref{subsec:2dHalfGauging}.
We split the lattice $\Gamma$ into two semi-infinite regions $\cA$ and $\cB$ such that
\begin{equation}
    \Gamma = \cA \sqcup \cB,
    \qquad
    \partial\cA = \cW_\cA,
    \qquad
    \partial\cB = \cW_\cB,
\end{equation}
and localize the interface on the joint boundary
\begin{equation}
    \cW = \cW_\cA\sqcup\cW_\cB.
\end{equation}
Region $\cA$ will carry the original $\Z_2^{(1)}$ description, while region $\cB$ is gauged.
We choose the boundary condition on $\cW_\cB$ to be ``rough'' (i.e.\ the interface does not terminate on vertices), so that we do not have to modify the Gauß laws close to the boundary.

\begin{figure}
    \centering
\includegraphics[width=0.8\linewidth]{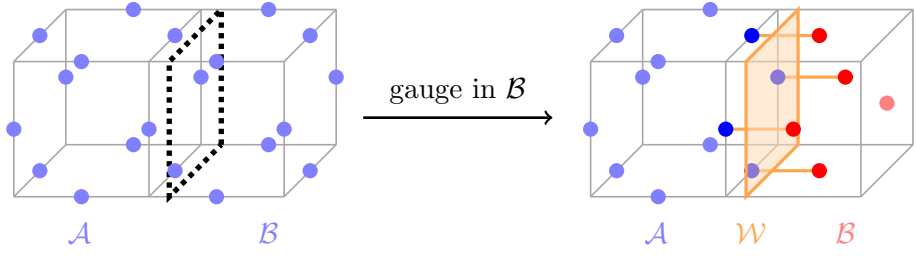}
    \caption{Graphical depiction of $3d$ half gauging. We gauge in sublattice $\cB$ such that the model hosts dual degrees of freedom on plaquette centers (depicted in red). The degrees of freedom participating in interactions across the boundary are shaded in a darker color. Together, they make up the interface, denoted by $\cW = \cW_A\sqcup \cW_B$. In this simple case of a flat interface, these interactions are all given by $Z_eX_p$ where $e,p$ is a pair of neighboring edge and plaquette. This is denoted by an orange line in the above figure.} 
    \label{fig:3dhalfgauging}
\end{figure}

As illustrated in Fig.~\ref{fig:3dhalfgauging}, gauging only in region $\cB$ leads to the interface Hamiltonian
\begin{equation}
    H_{\cD(\cW)} = H_{\cA} + H_{\cB} + H_{\cW}\,,
\end{equation}
with
\begin{align}
    \begin{split}
        H_{\cA} &= -\sum_{p\in\cA}B_p - \sum_{e\in\cA}X_e\,, \\
        H_{\cB} &= -\sum_{e\in\cB} A_e - \sum_{p\in \cB/\cW_\cB} X_p\,, \\
        H_{\cW} &= - \sum_{p\in\cW_\cB} X_p \left(\prod_{e\in p,e\in\cW_\cA}Z_e\right)\,.
    \end{split}
    \label{eq:1formInterfaceHamiltonian}
\end{align}
Away from $\cW$, $H_{\cA}$ reduces to the original $\Z_2$ lattice gauge theory~\eqref{eq:Z2LGT}, while $H_{\cB}$ is the dual model~\eqref{eq:3+1ddual} with plaquette variables.
One could further dualize only region $\cB$ to obtain a description purely in terms of edge qubits and study the duality defects in a single theory. Note, however, that this would obscure the locality of the cross-defect couplings. This is why we opt here to continue working with this mixed description.

\medskip\noindent
Similar to the case in $(2+1)d$, there is a unitarily equivalent description obtained by starting from the dual $\Z_2^{(1)}$ theory and gauging only in region $\cA$
\begin{equation}
    \widetilde H_{\cD(\cW)} = \widetilde H_\cA + \widetilde H_\cB + \widetilde H_\cW\,,
\end{equation}
with
\begin{align}
    \begin{split}
        \widetilde H_\cA &= -\sum_{p\in\cA} B_p - \sum_{e\in\cA/\cW_\cA}X_e\,, \\
        \widetilde H_\cB &= - \sum_{e\in\cB}A_e - \sum_{p\in\cB}X_p\,, \\
        \widetilde H_\cW &= -\sum_{e\in\cW_\cA}X_e \left(\prod_{p\ni e,p\in\cW_\cB}Z_p\right)\,.
    \end{split}
\end{align}
These two presentations are related by a unitary localized on the interface,
\begin{equation}
    \widetilde H_{\cD(\cW)} = U_\cW\, H_{\cD(\cW)}\, U_\cW^\dagger,
    \qquad
    U_\cW= \prod_{p\in\cW_\cB}\prod_{e\in (p\cap \cW_\cA)} \mathsf{CZ}_{e,p}\,.
\end{equation}
It will be useful to use different presentations on different sides when analyzing fusion and movement.

\paragraph{Fusion with Symmetry Defects.}
Let $\eta_1(\Sigma_1^\vee)$ denote a $\Z_2^{(1)}$ defect inserted on a closed, non-contractible string $\Sigma_1^\vee$ on the dual lattice.\footnote{If $\Sigma_1^\vee$ is contractible,we can use an appropriate product of $X_e$ to completely trivialize $\eta_1(\hat\Sigma_1)$.} The Hamiltonian with both a $\eta_1(\Sigma^\vee_1)$ defect and the duality interface $\cD(\cW)$ is
\begin{equation}
    H_{\eta(\Sigma_1^\vee),\cD(\cW)} = -\sum_{p\in\cA}(-1)^{\delta(\Sigma_1^\vee,p)}B_p - \sum_{e\in\cA}X_e -\sum_{e\in\cB} A_e - \sum_{p\in \cB/\cW_\cB} X_p - \sum_{p\in\cW_\cB} X_p \left(\prod_{e\in p,e\in\cW_\cA}Z_e\right)\,,
\end{equation}
where, without loss of generality, we have taken $\hat\Sigma_1\subset\cA$ and we have defined $\delta(\Sigma_1^\vee,p) = 1$ if $p\in \Sigma_1^\vee$ and $0$ otherwise. The local movement operator of $\eta_1$ is given by a single $X_e$. Then, we use the movement operators to move $\eta_1$ to the boundary
\begin{equation}
    H_{\eta(\Sigma_1^\vee),\cD(\cW)} \cong -\sum_{p\in\cA}B_p - \sum_{e\in\cA}X_e -\sum_{e\in\cB} A_e - \sum_{p\in \cB/\cW_\cB} X_p - \sum_{p\in\cW_\cB}(-1)^{\delta(\Sigma_1^\vee,p)} X_p \left(\prod_{e\in p,e\in\cW_\cA}Z_e\right)\,.
\end{equation}
Finally, acting with the unitary $U_{\cW_\cB}=\prod_{p\in\cW_\cB}Z_p$ brings us back to the original interface Hamiltonian. Therefore, the duality interface absorbs the $1$-form symmetry defect,
\begin{equation}
    \eta_1(\Sigma_1^\vee)\otimes \cD(\cW) \;\cong\; \cD(\cW),
    \qquad
    \cD(\cW)\otimes \eta_1(\Sigma_1^\vee) \;\cong\; \cD(\cW),
\end{equation}
in precise analogy with~\eqref{eq:etaDFusion} and~\eqref{eq:DetaFusion}.

\subsection{Movement Operator}
\label{subsec:1formmovement}
The defect $\cD$ is topological and can therefore be deformed and moved on the lattice by unitaries. We now make this explicit by constructing its (minimal) movement operator. We claim that, in general spatial dimension $d$, the movement operator for gauging $\Z_2^{(1)}$-symmetry is given by 
\begin{align}
    \begin{split}
    \lambda_{p_0,e_0}^{1,d-2} &= \left(\prod_{e\in (p_0\cap \cB),p\in(\cB/p_0),<e,p>} \mathsf{CZ}_{e,p}\right)\mathsf{S}_{e_0,p_0}\mathsf{H}_{p_0}\prod_{e\in (p_0\cap\cB/e_0)}\mathsf{CZ}_{e,p}\,, \\
    \widetilde\lambda_{p_0,e_0}^{1,d-2} &= \mathsf{S}_{e_0,p_0}\mathsf{H}_{p_0}\left(\prod_{e\in p_0,e\neq e_0}\mathsf{CZ}_{e,p}\right)\,.
    \end{split}
    \label{eq.MovementOp3d}
\end{align}
Here, $\lambda_{p_0,e_0}$ realizes an elementary move which removes a plaquette $p_0\in \cW_\cB$ and adds the appropriate edges degrees of freedom around it, thereby enlarging the region $\cA$ by the edges surrounding $p_0$.
\paragraph{Motivation from $(2+1)d$.}
To motivate the form of the movement operator, let us temporarily restrict to $d=2$, where we can relate the movement operator for the 1-form gauging \eqref{eq.MovementOp3d} to the 0-form case studied in Section~\ref{sec:0form}. We expect 
\begin{equation}
    \lambda^{1,0}_{p_0,e_0} \sim \left(\lambda^{0,1}_{e_0,v_0}\right)^{-1}\,,
\end{equation}
up to dualization and (possibly) an additional local unitary. Here $\lambda^{1,0}_{p_0,e_0}$ removes the plaquette $p_0$ from the 1-form region and introduces a degree of freedom on the edge $e_0$ (and possibly on further edges, depending on the constraints), while $\lambda^{0,1}_{e_0,v_0}$ was constructed in Section~\ref{subsec:0formmovement}. 

Next, observe that the interface in Eq.~\eqref{eq:1formInterfaceHamiltonian} is of the same form as the alternative description $\widetilde{H}_{\cW}$ in Eq.~\eqref{eq:0formAlternativeInterface}. Therefore, in analogy with Eq.~\eqref{eq:AlternativeMovement}, it is natural to expect that
\begin{align}
    \begin{split}
    \lambda_{p_0,e_0}^{1,0} &= T_{\bm{\frac{1}{2}}} \left(\widetilde\lambda^{0,1}_{e_0,v_0}\right)^{-1}=T_{\bm{\frac{1}{2}}} U_{\cW}\left(\lambda^{0,1}_{e_0,v_0}\right)^{-1}U_{\cW'}^\dagger  \\&= \left(\prod_{e\in (p_0\cap \cB),p\in(\cB/p_0),<e,p>} \mathsf{CZ}_{e,p}\right)\mathsf{S}_{e_0,p_0}\mathsf{H}_{p_0}\prod_{e\in (p_0\cap\cB/e_0)}\mathsf{CZ}_{e,p_0}\,.
    \end{split}
\end{align}
On the other hand, for $\widetilde{ \lambda}_{p_0}^{1,0}$, we expect the simpler expression
\begin{equation}
    \widetilde\lambda_{p_0,e_0}^{1,0} = T_{\bm{\frac{1}{2}}}\lambda_{e_0,v_0}^{-1}=\mathsf{S}_{e_0,p_0}\mathsf{H}_{p_0}\prod_{e\in p_0,e\neq e_0}\mathsf{CZ}_{e,p_0}\,.
\end{equation}

\begin{figure}
    \centering
    \includegraphics[width=0.8\linewidth]{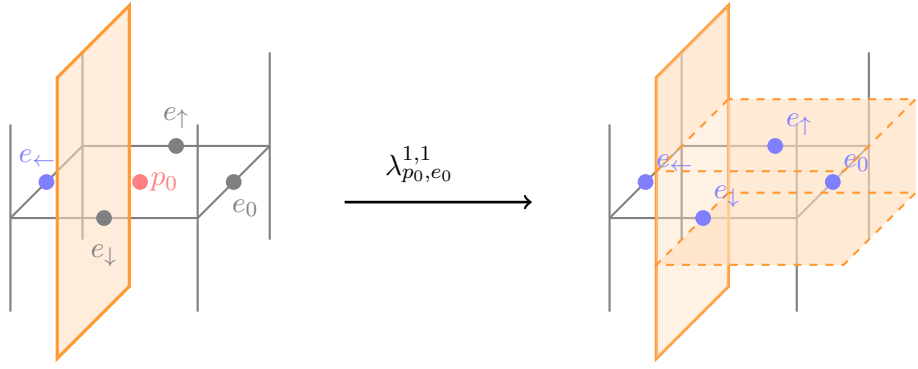}
    \caption{An exemplary movement in 3d which removes the plaquette $p_0$ from region $\cB$ and adds the surrounding edges $e_\uparrow$, $e_\downarrow$ and $e_0$.}
    \label{fig:LocalMovementOp3d}
\end{figure}

\paragraph{An explicit check.}
Let us work out a representative local move in $(3+1)d$. We consider the interface illustrated in Figure \ref{fig:LocalMovementOp3d} and calculate the elementary movement operator $\widetilde\lambda_{p_0,e_0}^{1,1}$. This removes the plaquette $p_0$ from region $\cB$ and introduces the edge $e_0$, together with the additional edge degrees of freedom required by the constraints.

The movement operator $\widetilde\lambda_{p_0,e_0}^{1,1}$ acts non-trivially on the following terms of $\widetilde H_{\cD(\cW)}$:
\begin{subequations}
    \begin{align}
        \widetilde\lambda_{p_0,e_0}^{1,1}:X_{e_\leftarrow}Z_{p_0} &\mapsto X_{e_\leftarrow} \\
        X_{p_0} &\mapsto B_{p_0} \\
        A_{e_0} &\mapsto X_{e_0}\prod_{p\ni e_0,p\neq p_0}Z_p \\
        A_{e_\downarrow} &\mapsto X_{e_0}\prod_{p\ni e_\downarrow,p\neq p_0}Z_p \label{eq:ConstraintNeeded1}\\
        A_{e_\uparrow} &\mapsto X_{e_0}\prod_{p\ni e_\uparrow,p\neq p_0}Z_p 
        \label{eq:ConstraintNeeded2}
    \end{align}
\end{subequations}
In addition, before the move we impose the local constraints 
\begin{equation}
    X_{e_\downarrow} = X_{e_\uparrow} = X_{e_0} = 1\,,
\end{equation}
so that the new edge degrees of freedom are initialized in the appropriate symmetric state.
After acting with $\widetilde\lambda_{p_0,e_0}^{1,1}$ these constraints become
\begin{subequations}
    \begin{align}
    \widetilde\lambda_{p_0,e_0}^{1,1}: X_{e_0} &\mapsto X_{p_0} = 1\label{eq:3dXp}\,,\\
    X_{e_\downarrow} &\mapsto X_{e_\downarrow}X_{e_0} = 1\label{eq:3dXedown} \,,\\
    X_{e_\uparrow} &\mapsto X_{e_\uparrow}X_{e_0} =1\label{eq:3dXeup}\,,
    \end{align}
\end{subequations}
which imply $X_{p_0}=X_{e_\downarrow}=X_{e_\uparrow}$~\footnote{Note that these constraints only hold at this step of the movement. After subsequent movements, they will eventually become $\prod_{e\ni v}X_e = 1$, i.e. they ensure that the symmetry is topological.} and freeze the plaquette degree of freedom at $p_0$.
The net effect of the move is thus to trade the plaquette degree of freedom for edge degrees of freedom, compatibly with the Gauß constraints and the interface terms. Equivalently after using the constraints we can rewrite
\begin{subequations}
    \begin{align}
        A_{e_\downarrow} &\mapsto X_{e_\downarrow}\prod_{p\ni e_\downarrow,p\neq p_0}Z_p\\
        A_{e_\uparrow} &\mapsto X_{e_\uparrow}\prod_{p\ni e_\uparrow,p\neq p_0}Z_p \,,
    \end{align}
\end{subequations}
Collecting everything, we obtain
\begin{equation}
    \widetilde\lambda^{1,1}_{p_0,e_0}\,
    \widetilde H_{\cD(\cW)}\,P(\cW)\,
    \big(\widetilde\lambda^{1,1}_{p_0,e_0}\big)^\dagger
    = 
    \widetilde H_{\cD(\cW')}\,P(\cW')\,,
    \label{eq:3dMovement}
\end{equation}
where $P(\cW)$ and $P(\cW')$ project onto the corresponding constrained Hilbert spaces. All other elementary moves follow from analogous calculations.
By composing such elementary moves along a path, we can transport the interface $\cD(\cW)$ to any other locus $\cD(\cW')$, making its topological nature manifest.

\subsection{Condensation Defects}
Again, we define the condensation defects $\cC_1(\Sigma)$ by gauging $\Z_2^{(1)}$ on a sublattice $\Sigma$, i.e.
\begin{equation}
    \cC_1(\Sigma) \cong \cD(\partial\Sigma)
\end{equation}
As before, the structure of $\cC_1(\Sigma)$ depends on the number of non-contractible cycles of $\Sigma$. In general, the Hamiltonian with a 1-form condensation defect reads
\begin{equation}
    H_{\cC_1(\Sigma)} = -\sum_{p\notin\Sigma}B_p - \sum_{e\notin\Sigma} X_e - \sum_{p\in(\Sigma/\partial\Sigma)}X_p - \sum_{e\in\Sigma}A_e-\sum_{p\in\partial\Sigma}X_p\left(\prod_{e\in p,e\notin\Sigma}Z_e\right)
\end{equation}
where $\partial\Sigma\subset\Sigma$ is the boundary of $\Sigma$.

\paragraph{Contractible $\Sigma$.} If $\Sigma$ is fully contractible, we can use the movement operator \eqref{eq.MovementOp3d} to deform $\Sigma\cong \Sigma^\vee_0=p_0$, with $\Sigma^\vee_0$ a single plaquette $p_0$. The resulting Hamiltonian is
\begin{equation}
    H_{\cC_1(\Sigma)} \cong H_{\cC_1(p_0)} = -\sum_{p\neq p_0}B_p -\sum_e X_e - X_{p_0}B_{p_0}
\end{equation}
Before inserting the condensation defect, we imposed the operator identity $\prod_{p\in S_2}B_p = 1$ for any closed, contractible two-dimensional sublattice $S^2$. For $\Sigma_0\in S^2$, this becomes $X_{\Sigma_0} =1$, so that $\cC_1(\Sigma_0) \cong 1$.

\paragraph{A single non-contractible cycle.} Let $\Sigma$ be non-contractible and wrap a single cycle $\Sigma_1^\vee$ on the dual lattice.
Then, the Hamiltonian reads
\begin{equation}
    H_{\mathcal C_1(\Sigma_1^\vee)} =  -\sum_{p\notin\Sigma_1^\vee}B_p - \sum_{e} X_e -  \sum_{p\in\Sigma_1^\vee}X_pB_p\,.
\end{equation}
We again impose the additional constraint $\prod_{p\in (S^2\cap \Sigma_1)}X_p =1$ for arbitrary closed, contractible 2-sublattices $S^2$. Since the lattice is three-dimensional, $\Sigma^\vee_1$ intersects $\Sigma^2$ an even number of times, so the constraint reduces to $X_{p} = \pm 1$ for all $p\in\Sigma_1^\vee$. For the defect, this implies
\begin{equation}
    H_{\mathcal C_1(\Sigma_1)} \cong H_1 \oplus H_{\eta_1(\Sigma_1)}\,,
\end{equation}
i.e., $\cC_1(\Sigma_1) \cong 1\oplus \eta_1(\Sigma_1)$.

\paragraph{Two non-contractible cycles.} Now, consider the defect on a two-dimensional sublattice $\Sigma_2$ with two non-trivial cycles. The Hamiltonian becomes
\begin{equation}
    H_{\mathcal C_1(\Sigma_2)} = -\sum_{p\notin\Sigma_2}B_p - \sum_{e\notin\Sigma_2} X_e - \sum_{e\in\Sigma_2}A_e-\sum_{p\in\Sigma_2}X_p\left(\prod_{e\ni p,e\notin\Sigma_2}Z_e\right)
\end{equation}
together with the new constraint $\prod_{p\in \Sigma_2\cap S^2}X_p = 1$ for all closed, contractible surfaces $S^2$. This is equivalent to demanding
\begin{equation}
    \prod_{p\in (\Sigma_2\cap c) } X_p =1
\end{equation}
where $c$ is any 3-cell of the lattice. This operator generates a topological $\Z_2^{(1)}$-symmetry on $\Sigma_2$. Therefore, the condensation defect $\mathcal{C}_1(\Sigma_2)$ realizes a $(2+1)d$ $\Z_2$ gauge theory on $\Sigma_2$, which is coupled to the bulk through the term $\sum_{p\in\Sigma_2}X_p\left(\prod_{e\ni p,e\notin\Sigma_2}Z_e\right)$.

\paragraph{Fusion with invertible defects.} Recall that in $(2+1)d$, when the non-trivial $\mathcal{C}_1$ condensation defect absorbs a bulk $\Z_2^{(1)}$ defect, it creates a local charge on the condensation defect. Here, we will show the analogous $(3+1)d$ statement. The bulk $\Z_2^{(1)}$ defect can be absorbed by the condensation defect $\cC_1(\Sigma_2)$ at the expense of creating a one dimensional Wilson line on $\Sigma_2$.

\begin{figure}
    \centering
    \includegraphics[width=0.55\linewidth]{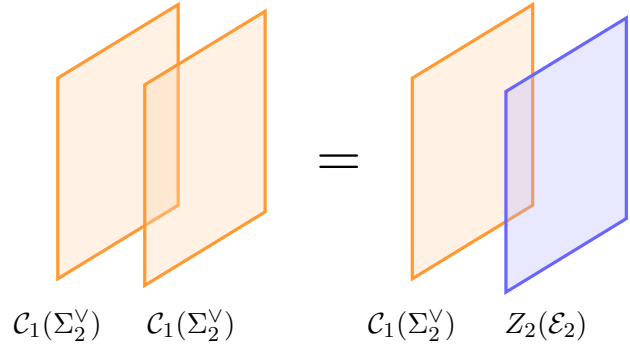}
    \caption{Fusing two condensation defects produces a single condensation defect and a 2d $\Z_2$-gauge theory. }
    \label{fig:C1C1Fusion3d}
\end{figure}

Let $\eta_1(\Sigma_1^\vee)$ be a bulk $\Z_2^{(1)}$ defect. We move it onto the condensation defect on $\widetilde\Sigma_1^\vee\subset\Sigma_2$ by the usual $\Z_2^{(1)}$ movement operator. This flips the sign of a collection of plaquette terms in the Hamiltonian terms as
\begin{equation}
    -\sum_{p\in\widetilde\Sigma_1^\vee}X_p\left(\prod_{e\in p,e\notin\Sigma_2}Z_e\right) \longrightarrow +\sum_{p\in\widetilde\Sigma_1^\vee}X_p\left(\prod_{e\in p,e\notin\Sigma_2}Z_e\right)\,.
\end{equation}
One can revert back to the original condensation defect Hamiltonian by applying $W_e(\Sigma^\vee_1)=\prod_{p\in \Sigma^\vee_1} Z_p$. Note that $W_e(\Sigma^\vee_1)$ is the electric Wilson line of the 2d $\Z_2$ gauge theory localized on $\Sigma_2$. If $\Sigma^\vee_1$ has endpoints, it creates charged particles on the endpoints. If $\Sigma^\vee_1$ is a contractible loop, then it acts trivially. If $\Sigma^\vee_1$ is a non-contractible loop, $W_e(\Sigma^\vee_1)$ changes the topological sector of the $(2+1)d$ gauge theory realized on the condensation defect.
\begin{figure}
    \centering
    \includegraphics[width=0.75\linewidth]{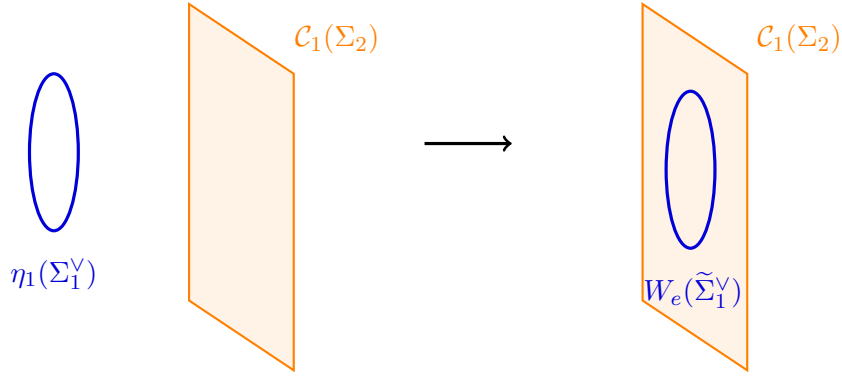}
    \caption{The bulk $\mathbb{Z}_2^{(1)}$defect $ \eta_1(\Sigma^{\vee}_1)$ can be moved onto the condensation defect $\cC_1(\Sigma_2)$, where it becomes a Wilson line $W_e(\Sigma_1^{\vee})$ of the $\Z_2$ gauge theory defined on $\Sigma_2$.}
    \label{fig:conddefect3d}
\end{figure}
\paragraph{Fusion of two condensation defects.} Next, we discuss the fusion of two condensation defects. The Hamiltonian with two condensation defects supported on two surfaces on the dual lattice $\Sigma^\vee_2$ and $\widetilde\Sigma_2^\vee$ reads
\begin{align}
    \begin{split}
        H_{\cC_1(\Sigma_2),\cC_1(\Sigma_2')} =& -\sum_{p\notin (\Sigma_2^\vee\cup\widetilde\Sigma^\vee_2)}B_p - \sum_{e\notin (\Sigma_2^\vee\cup\widetilde\Sigma_2^\vee)}X_e - \sum_{e\in \Sigma_2^\vee\cup\widetilde\Sigma_2^\vee}A_e \\ &- \sum_{p\in \Sigma_2^\vee}\left(\prod_{e\in p,e\notin\Sigma_2^\vee}Z_e\right)X_p - \sum_{p\in \widetilde\Sigma_2^\vee}\left(\prod_{e\in p,e\notin\widetilde\Sigma_2^\vee}Z_e\right)X_p\,.
    \end{split}
\end{align}
We now apply movement operators to bring the defects into the geometry shown in Figure~\ref{fig:C1C1Fusion3d}: $\Sigma_2^\vee$ and $\widetilde\Sigma_2^\vee$ are now taken to be flat surfaces on the dual lattice adjacent to one another. In between lies a surface on the direct lattice. Let us denote this surface by $\mathcal{E}_2$. We also define $\mathcal{S}=\Sigma_2^\vee\cup\mathcal{E}_2\cup\widetilde\Sigma_2^\vee$. Note that it holds that the boundary of $\mathcal{S}$ is given by $\partial\mathcal{S} = \Sigma_2\cup \Sigma_2'$. Then, upon applying the fusion unitary
\begin{equation}
    \prod_{p\in \partial\mathcal{S}}\left( \prod_{e\in p,e\notin\partial\mathcal{S}}\mathsf{CZ}_{e,p}\right)\,,
\end{equation}
the Hamiltonian becomes
\begin{align}
    \begin{split}
        H_{\cC_1(\Sigma_2),\cC_1(\Sigma_2')} &\cong -\sum_{p\notin \mathcal{S}}B_p - \sum_{e\notin \mathcal{S}}X_e - \sum_{e\in\mathcal{S}/\mathcal{E}_2}A_e -\sum_{p\in \partial\mathcal{S}}\left(\prod_{e\in p,e\notin\mathcal{S}}Z_e\right)X_p  \\
        &\quad-\sum_{p\in\mathcal{E}_2}B_p - \sum_{e\in\mathcal{E}_2}\left(\prod_{p\ni e,p\notin\mathcal{E}_2}Z_p\right)X_e\\
        &=H_{\mathcal{C}_1(\mathcal{S}),\mathcal{Z}(\mathcal{E}_2)}\,.
    \end{split}
\end{align}
We notice that on $\mathcal{E}_2$, the Hamiltonian consists of plaquette terms $B_p$ and transverse-field terms $X_e$, which are coupled to the bulk. Therefore, we can identify this with a 2d $\Z_2$ gauge theory on $\mathcal{E}_2\subset \mathcal{S}$, which we denote by $\mathcal{Z}_2(\mathcal{E}_2)$. The rest of the Hamiltonian has the form of a single condensation defect on $\mathcal{S}$.
Consequently, we find the fusion rule
\begin{equation}
    \cC_1(\Sigma_2^\vee)\otimes\cC_1(\widetilde \Sigma_2^\vee)\cong \mathcal{Z}_2(\mathcal{E}_2)\otimes \cC_1(\mathcal{S})\,,
\end{equation}
where $\mathcal{Z}_2$ denotes a $(2+1)d$ $\Z_2$ lattice gauge theory. We find that, unlike in $(2+1)d$, the fusion coefficients in $(3+1)d$ may not be numbers but instead they form their own non-trivial physical theories defined on codimension one submanifolds.

\medskip\noindent
To conclude, the self-dual $\Z_2^{(1)}$-symmetric Hamiltonian admits the following non-trivial topological defects on the lattice: $\eta_1(\Sigma_1)$, $\cD(\Sigma_2)$, $\cC_1(\Sigma_2)$. Suppressing the explicit defect loci, the fusions of topological defects for a $\Z_2^{(1)}$-theory in $(3+1)d$ can be summarized as
\begin{subequations}
    \begin{align}
        \eta_1 \otimes \eta_1 &\cong 1\,, \\
        \eta_1 \otimes \cD& \cong  \cD\,, \\
        \cD \otimes \cD &\cong \cC_1 \,,\\
        \cD \otimes \cC_1 &\cong \cC_1 \,,\\
        \cC_1 \otimes \cC_1 &\cong \mathcal{Z}_2 \cC_1 \,.
    \end{align}
\end{subequations}
These same fusion rules have been derived in continuum field theory~\cite{Roumpedakis2023,Choi2023,Cordova:2024mqg}. Here, we confirm that the lattice realization realizes the same fusion rules.

\section{Application to Gauging Subsymmetries}
\label{sec:Subsym}
In the previous sections we constructed lattice gauging maps and topological interfaces for $\Z_2^{(p)}$ symmetries in various spacetime dimensions. 
The same construction extends straightforwardly to $\Z_n^{(p)}$ for arbitrary $n$, using clock and shift operators and their generalized controlled gates.
However, for $n\ge 4$ a qualitatively new phenomenon appears: one can \emph{gauge only a subgroup} $\Z_m^{(p)}\subset \Z_n^{(p)}$, and the resulting theory then carries a non-trivial higher-group symmetry
\begin{equation}
    G^\epsilon_{p,d-p-1} = [\Z_{n/m}^{(p)},\Z_{m}^{(d-p-1)}]^\epsilon
\end{equation}
with a mixed 't~Hooft anomaly $\epsilon$~\cite{Tachikawa:2017gyf,Bhardwaj2023}.
In this section we study this situation on the lattice, focusing on the simplest non-trivial example of gauging a $\Z_2$ subgroup of a $\Z_4$ symmetry, in arbitrary spatial dimension $d$.
We begin by reviewing the construction of the corresponding gauged Hamiltonian by gauging a subsymmetry, as was first explored in~\cite{Moradi2025}. We then perform sequential gauging, analyze the corresponding interfaces and defects, and symmetry fractionalization, thereby confirming a number of categorical results~\cite{Bhardwaj2023} explicitly on the lattice.

\subsection{Anomalies and Defects from Subgroup Gauging}\label{subsec:subgauging}

\paragraph{\texorpdfstring{$\Z_4^{(0)}$}{Z4(0)} models.}
We start from a $\Z_4^{(0)}$-symmetric lattice model on a $d$-dimensional spatial lattice $\Gamma_d$ with Hilbert space $
\mathcal{H} = \bigotimes_{v\in\Gamma_d} \C^4_v$ on the vertex.
To each vertex $v$ we assign a $\Z_4$ qudit $\ket{s_v}\in \C^4_v$ with clock and shift operators $Z_v$ and $X_v$ acting as
\begin{equation}
    Z_j\ket{s_j} = \omega^{s_j}\ket{s_j},\quad 
    X_j\ket{s_j} = \ket{s_j+1},\quad  \omega=\exp(2\pi i/4)=i\,.
    \label{eq:ClockShift2}
\end{equation}
One canonical example is the $\Z_4$ qudit TFIM
\begin{equation}\label{eq:z4tfim}
    H_{\Z^{(0)}_4} = -\sum_{e} Z_{s(e)}^\dagger Z_{t(e)} -\sum_v X_v + \text{h.c.}\,,
\end{equation}
where the sum runs over \textit{oriented} edges $e$ of $\Gamma_d$, with source and target vertices $s(e)$ and $t(e)$.
The onsite global $\Z_4^{(0)}$ symmetry is generated by
\begin{equation}
    U = \prod_{v} X_v\,,
\end{equation}
while the twisted sectors are labeled by $\Z_4$-valued twist operators
\begin{equation}
    V(\Sigma_1) = \prod_{e\in\Sigma_1} Z_{s(e)}^\dagger Z_{t(e)}\,,
\end{equation}
for closed one-dimensional  sublattices $\Sigma_1$.
We require $V(\Sigma_1)=1$ for all contractible $\Sigma_1$.
\paragraph{Gauging \texorpdfstring{$\Z_2^{(0)}\subset\Z_4^{(0)}$}{Z2(0) in Z4(0)}.}
Let us now review the gauging of the $\Z_2^{(0)}$-subgroup generated by $U^2$~\cite{Moradi2025}. We introduce $\Z_2$ gauge degrees of freedom on edges,
\begin{equation}
    \mathcal{H}_{\text{ext}} = 
    \Big(\bigotimes_{v\in\Gamma_d} \C^4_v\Big)\otimes
    \Big(\bigotimes_{e\in\Gamma_d} \C^2_e\Big),
\end{equation}
with Pauli operators $\sigma_e^x,\sigma_e^z$ acting on edge $e$.
After the minimal coupling 
$Z_{s(e)}^\dagger Z_{t(e)} \to Z_{s(e)}^\dagger \,\sigma_e^x\, Z_{t(e)}$, the gauged Hamiltonian is invariant under local $\Z_2$ gauge transformations generated by the Gauß law on each vertex
\begin{equation}
    G_v = X_v^2 A_v,\quad
    A_v = \prod_{e\ni v}\sigma_e^z\,,
\end{equation}
and physical states obey $G_v=1$ for all vertices $v$.
Similar to the $\Z_2$ case, we use the unitary
\begin{equation}
    \mathcal{M} = \prod_{v}\prod_{e\ni v} \widetilde{\mathsf{CZ}}_{v,e}\,
\end{equation}
where 
\begin{equation}    
    \widetilde{\mathsf{CZ}}_{v,e} 
    = \frac{1}{4}\sum_{\alpha=1}^4\sum_{\beta=1}^4 
       \omega^{-\alpha\beta} Z_v^\alpha (\sigma_e^z)^{\lfloor\beta/2\rfloor}\,,
\end{equation}
is the mixed $\Z_4$–$\Z_2$ generalization of the controlled-$Z$ gate. Therefore, $\mathcal{M}$ localizes the Gauß operator to a single on-site operator
\begin{equation}
    \mathcal{M}_v G_v \mathcal{M}_v^\dagger = X_v^2\,,
\end{equation}
such that the Gauß law $X_v^2=1$ can be imposed directly.
After projection onto the gauge invariant sector with ``reduced'' Hilbert space
\begin{equation}
    \mathcal{H}_{\text{rest}} = 
    \Big(\bigotimes_{v}\text{span}\{\ket{0_v}+\ket{2_v},\ket{1_v}+\ket{3_v}\}\Big)
    \otimes\Big(\bigotimes_e \C_e^2\Big)
    \cong \Big(\bigotimes_{v}\C_v^2\Big)\otimes\Big(\bigotimes_e\C_e^2\Big)\,,
\end{equation} 
we can define effective Pauli operators
\begin{equation}
    Z_v^2\big|_{X_v^2=1} = \sigma_v^z\,,\qquad
    X_v\big|_{X_v^2=1} = \sigma_v^x\,.
    \label{eq:effetivedof}
\end{equation}
Employing the transformation of local operators under conjugation of unitary $\mathcal{M}$
\begin{align}
    \mathcal{M} X_v \mathcal{M}^\dagger &= X_v (P_v^+ + P_v^- A_v),\quad P_v^\pm = \frac{1}{2}(1\pm Z_v^2)\,,\\
    \mathcal{M} Z^{\dagger}_{s(e)}\sigma_e^x Z_{t(e)} \mathcal{M}^\dagger 
    &= \frac{1}{2}(1 - i Z^2_{s(e)}) \,\sigma_e^x\, (1+i Z^2_{t(e)})\,,
\end{align}
together with these effective variables~\eqref{eq:effetivedof}, we obtain the gauged models. For example, starting from the $\Z_4$ qudit TFIM~\eqref{eq:z4tfim}, the gauged Hamiltonian is
\begin{align}
    \begin{split}
    H_{\Z_2^{(0)}\times\Z_2^{(d-1)}} &= -\sum_e\frac{1}{2}(1-i\sigma^z_{s(e)})\sigma_e^x(1+i\sigma^z_{t(e)})-\sum_v\sigma_v^x(P_v^++P_v^-A_v) + \text{h.c.}
    \label{eq:HZ2Z2} \\
    &= -\sum_e \sigma_e^x(1+\sigma_{s(e)}^z\sigma_{t(e)}^z) - \sum_v \sigma_v^x((1+\sigma_v^z)+(1-\sigma_v^z)A_v)\,.
    \end{split}
\end{align}
\paragraph{Mixed anomaly and symmetry fractionalization.}
The global symmetry of the gauged theory is generated by 
\begin{subequations}
    \begin{align}
        \widetilde{U}&=\mathcal{M}U \mathcal{M}^\dagger|_{X_v^2=1} = \prod_v\sigma^x_v(P_v^++P_v^-A_v) = \prod_v\widetilde{u}_v \,, \\
        \widetilde{V}(\Sigma_1)&=\mathcal{M}V\prod_e \sigma_e^x \mathcal{M}^\dagger|_{X_v^2=1}=\prod_{e\in\Sigma_1}\left[\frac{1}{2}(1-i\sigma^z_{s(e)})\sigma_e^x(1+i\sigma^z_{t(e)})\right] = \prod_{e\in\Sigma_1}\widetilde{v}_e\,,
    \end{align}
\end{subequations}
with the following action on local operators
\begin{align}
\begin{aligned}
    \widetilde{U}: \quad 
    &\sigma_v^z \mapsto -\sigma_v^z,\quad 
     \sigma_e^z \mapsto \sigma_e^z,\\
    &\sigma_v^x \mapsto A_v\sigma_v^x,\quad
     \sigma_e^x \mapsto \sigma_{s(e)}^z\sigma_e^x\sigma_{t(e)}^z,
\end{aligned}
\quad
\begin{aligned}
    \widetilde{V}(\Sigma_1):\quad
    &\sigma_v^z \mapsto \sigma_v^z,\quad
     \sigma_e^z \mapsto -\sigma_e^z ,\\
    &\sigma_v^x \mapsto \sigma_v^x,\quad
     \sigma_e^x \mapsto \sigma_e^x \quad \forall v,e\in\Sigma_1\,.
\end{aligned}
\end{align}
Therefore, $\widetilde{U}$ generates a global $\Z_2^{(0)}$ symmetry, and $\widetilde{V}(\Sigma_1)$ generates a higher-form $\Z_2^{(d-1)}$ symmetry, together with a mixed 't~Hooft anomaly. This mixed anomaly is manifested by  
\begin{equation}
    \widetilde{U}^2 = \prod_vA_v,\quad\widetilde{V}(\Sigma_1)^2 = \prod_{e\in\Sigma_1}\sigma_{s(e)}^z\sigma_{t(e)}^z\,,
\end{equation}
where the square of each symmetry operator is the twist operator of the other symmetry~\cite{Moradi2025}.

\paragraph{$\Z_2$ Defects.}
Here, we construct the $\Z_2$ symmetry defects of the $\Z_2^{(0)}\times_{\epsilon}\Z_2^{(d-1)}$-theory, utilizing our framework of acting with "half" of a symmetry operators. Denote by $\widetilde \cU$ and $\widetilde \cV$ the defects corresponding to the symmetry operators $\widetilde{U}$ and $\widetilde{V}$, respectively. These defects will become important in Section \ref{subsubsec:SubSymFrac} as a new defect approach to diagnose symmetry fractionalization.
For the $\Z_2^{(0)}$-symmetry generated by $\widetilde U$, we split the lattice into two regions $\cA$ and $\cB$, with shared boundary $\cW$, and apply the symmetry operator only in region $\cB$. We find
\begin{align}
    \begin{split}
        H_{\widetilde\cU(\cW)} &= -\sum_v \sigma_v^x(P_v^++P_v^-A_v) - \frac{1}{2}\sum_{e\notin\cW_A}(1-i\sigma_{s(e)}^z)\sigma_e^x(1+i\sigma_{t(e)}^z) \\
        &\qquad-\frac{i}{2}\sum_{e\in\cW_A}(1-i\sigma_{s(e)}^z)\sigma_e^x(1+i\sigma_{t(e)}^z) + \text{h.c} \\
        &= - \sum_v \sigma_v^x((1+\sigma_v^z)+(1-\sigma_v^z)A_v) -\sum_{e\notin\mathcal{W}_A} \sigma_e^x(1+\sigma_{s(e)}^z\sigma_{t(e)}^z) -\sum_{e\in\mathcal{W}_A}\sigma_e^x(\sigma_{s(e)}^z-\sigma_{t(e)}^z)\,,
        \label{eq:SubSymDefectU}
    \end{split}
\end{align}
and for the $\Z_2^{(d-1)}$-symmetry we act with the one-dimensional operator $\widetilde V(\Sigma_1)$, defined on some semi-infinite string with endpoint $v_0$. Then, the Hamiltonian with a point-defect $\widetilde\cV(v_0)$ reads 
\begin{align}
    \begin{split}
        H_{\widetilde\cV(v_0)} = -\sum_{v\neq v_0}\sigma_v^x(P_v^++P_v^-A_v) - \frac{1}{2}\sum_{e}(1-i\sigma_{s(e)}^z)\sigma_e^x(1+i\sigma_{t(e)}^z) - \sigma_{v_0}^y(P_{v_0}^+-P_{v_0}^-A_{v_0}) + \text{h.c.}\,.
        \label{eq:SubSymDefectV}
    \end{split}
\end{align}
Since both these symmetries are group like, the movement operator is rather trivially given by the local symmetry operators $\widetilde u_v$ and $\widetilde v_e$, respectively. 

\subsection{Sequential Gauging on the Lattice}
\label{subsec:SequentialGauging}
Because of the mixed 't Hooft anomaly in $\Z_2^{(0)}\times_\epsilon\Z_2^{(d-1)}$, we can not gauge the full symmetry. We can, however, gauge either of the $\Z_2$ subgroups, as they are non-anomalous by themselves. Using a categorical argument~\cite{Bhardwaj2023} it has been argued that gauging the $\Z_2^{(0)}$ subgroup produces a dual theory with $\Z_4^{(d-1)}$ symmetry, whereas gauging the $\Z_2^{(d-1)}$ subgroup gets us back to the original theory with $\Z_4^{(0)}$ symmetry. In the following, we realize this sequential gauging explicitly on the lattice.

\paragraph{Gauging $\Z_2^{(d-1)}\subset \Z_2^{(0)}\times_\epsilon \Z_2^{(d-1)}$.} 
We first gauge the $\Z_2^{(d-1)}$ symmetry generated by $\widetilde{\cV}(\Sigma_1)$. Because this symmetry locally acts as
\begin{equation}
    \widetilde{v}_e: A_v\mapsto -A_v \quad \text{for }v\in e\,,
\end{equation}
we minimally couple to a $\Z_2$ gauge field $\tau_v^x$ on each vertex such that
\begin{equation}
    H_{(\Z_2^{(0)}\times\Z_2^{(d-1)})/\Z_2^{(d-1)}} = -\sum_v \sigma_v^x(P_v^++P_v^-\tau_v^xA_v)-\sum_e \sigma_e^x(1+\sigma_{s(e)}^z\sigma_{t(e)}^z)\,.
\end{equation}
The gauge symmetry is generated by the Gauß operator $G_e = \tau^z_{s(e)}\sigma_e^x\tau^z_{t(e)}$ on every edge.
Using the unitary $\mathcal{M}=\prod_{e}\mathsf{CZ}_{s(e),e}\mathsf{CZ}_{t(e),e}$, where $\mathsf{CZ}$ acts between the Hilbert spaces acted upon by $\tau_v$ and $\sigma_e$ respectively, we can localize the Gauß constraint 
\begin{equation}
    \mathcal{M}G_e\mathcal{M}^{\dagger}=\sigma^x_e=1\,,
\end{equation}
and project onto the gauge invariant sector. The resulting Hamiltonian is
\begin{equation}
    H_{(\Z_2^{(0)}\times\Z_2^{(d-1)})/\Z_2^{(d-1)}} = -\sum_v\sigma_v^x(P_v^++P_v^-\tau_v^x)-\sum_e \tau_{s(e)}^z\tau_{t(e)}^z(1+\sigma_{s(e)}^z\sigma_{t(e)}^z)\,,
\end{equation}
on the gauged Hilbert space $\cH = \bigotimes_v (\C^2\otimes \C^2)_v$.  With the isomorphism $ (\C^2\otimes \C^2)_v \cong  \C_v^4$ and the following transformation of basis states
\begin{equation}
    \ket{00}_v\mapsto \ket{0}_v \,,\quad\ket{01}_v\mapsto \ket{1}_v\,,\quad\ket{10}_v\mapsto \ket{2}_v\,,\quad\ket{11}_v\mapsto \ket{3}_v\,,
\end{equation}
we can rewrite the $\mathbb Z_2$ Pauli operators in terms of $\mathbb Z_4$ operators
\begin{align}
    \begin{split}
        \sigma_v^x(P_v^++P_v^-\tau_v^x) &\mapsto X_v + X_v^\dagger, \\
        \tau_{s(e)}^z\tau_{t(e)}^z(1+\sigma_{s(e)}^z\sigma_{t(e)}^z)&\mapsto Z_{s(e)}Z_{t(e)}^\dagger + Z_{s(e)}^\dagger Z_{t(e)}\,,
    \end{split}
\end{align}
and get back to the original  $\Z_4$-symmetric Hamiltonian~\eqref{eq:z4tfim}.

\paragraph{Gauging $\Z_2^{(0)}\subset \Z_2^{(0)}\times_\epsilon \Z_2^{(d-1)}$.}
The $\Z_2^{(0)}$-symmetry is generated by
\begin{equation}
    \widetilde{\cU}= \prod_v\sigma^x_v(P_v^++P_v^-A_v) = \prod_v u_v\,.
\end{equation}
Locally, it acts as
\begin{subequations}
\begin{align}
    \widetilde u_{s(e)}&: (1-i\sigma^z_{s(e)})\sigma_e^x(1+i\sigma^z_{t(e)})\mapsto -i(1-i\sigma^z_{s(e)})\sigma_e^x(1+i\sigma^z_{t(e)})\,, \\
    \widetilde u_{t(e)}&:(1-i\sigma^z_{s(e)})\sigma_e^x(1+i\sigma^z_{t(e)})\mapsto i(1-i\sigma^z_{s(e)})\sigma_e^x(1+i\sigma^z_{t(e)})\,.
\end{align}
\end{subequations}
The other term in the Hamiltonian \eqref{eq:HZ2Z2} is clearly left invariant. Notice that, locally, $\widetilde u_v$ acts like a $\Z_4$-symmetry. Therefore, we can make the Hamiltonian \eqref{eq:HZ2Z2} gauge-invariant by extending the Hilbert space to $\otimes_v \C_v^2 \otimes_e(\C^2\otimes \C^4)_e$ and minimally coupling
\begin{equation}
    (1-i\sigma^z_{s(e)})\sigma_e^x(1+i\sigma^z_{t(e)}) \mapsto (1-i\sigma^z_{s(e)})X_e\sigma_e^x(1+i\sigma^z_{t(e)})\,,
\end{equation}
where $X_e$ is a $\Z_4$ degree of freedom transforming according to $Z_e X_e=iX_eZ_e$ and the Hamiltonian acting on this extended Hilbert space now reads
\begin{equation}
    H= -\sum_e\frac{1}{2}(1-i\sigma^z_{s(e)})\sigma_e^xX_e(1+i\sigma^z_{t(e)})-\sum_v\sigma_v^x(P_v^++P_v^-A_v) + \text{h.c.}\,.
\end{equation}
The Gauß law at each site $v$ reads
\begin{equation}
    G_v = u_v \prod_{e\ni v}Z_e^{\sigma(v,e)} =1\,.
\end{equation}
where $\sigma(v,e)=\pm 1$, depending on whether the edge is coming in/out of the vertex.
Because $G_v$ has four eigenvalues, $\pm 1,\pm i$ and $G_v^2 = A_v\left(\prod_{e\ni v}Z_e^{(\dagger)}\right)^2 =1$ gives non-trivial constraints on the $\C^2_e$-degrees of freedom, the Hilbert space after imposing the Gauß law should reduce to $[\otimes_v \C_v^2 \otimes_e(\C^2\otimes \C^4)_e]_{G_v=1}\cong \otimes_e\C_e^4$.
We then define effective $\C_e^4$ Pauli operators $\widetilde{X}_e,\widetilde{Z}_e$ as
\begin{align}
    \begin{split}
      \widetilde{Z}_e = Z_e,\quad   \widetilde{X}_e = \frac{1}{2}(1-i\sigma^z_{s(e)})\sigma_e^xX_e(1+i\sigma^z_{t(e)})\,.
    \end{split}
\end{align}
Together with the constraint  $u_v = \prod_{e\ni v}\widetilde{Z}_e^{(\dagger)} = \widetilde{A}_v$ written in the effective operators, the final Hamiltonian 
\begin{equation}
    H = -\sum_v \widetilde{A}_v - \sum_e \widetilde{X}_e\,
\end{equation}
exhibits a $\Z_4^{(d-1)}$ symmetry.

\subsection{New Defect Diagnosis of Symmetry Fractionalization}
\label{subsubsec:SubSymFrac}
\begin{figure}
    \centering
    \includegraphics[width=\linewidth]{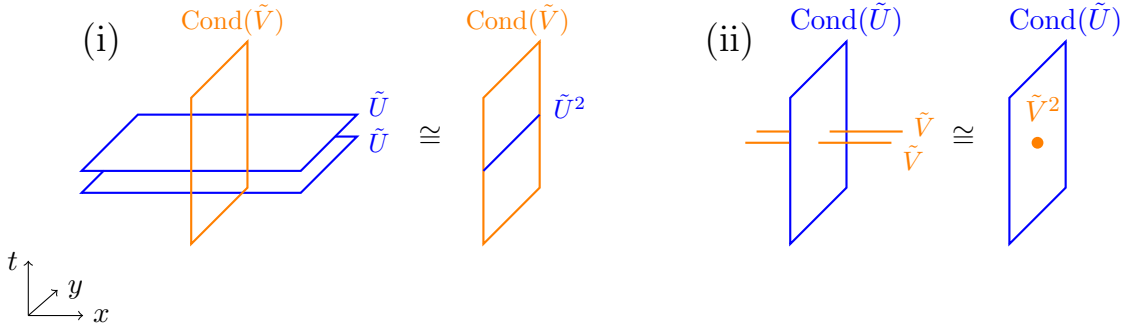}
    \caption{In (i), we fuse two $\widetilde U$ operators in the presence of a $\widetilde V$ condensation defect. In (ii), two $\widetilde V$ operators are fused in the presence of a $\widetilde U$ condensation defect. In both cases, the symmetry operator fractionalizes on the condensation defect.
    }
\label{fig:SymFracOnCond}
\end{figure}

\begin{figure}
    \centering
    \includegraphics[width=0.9\linewidth]{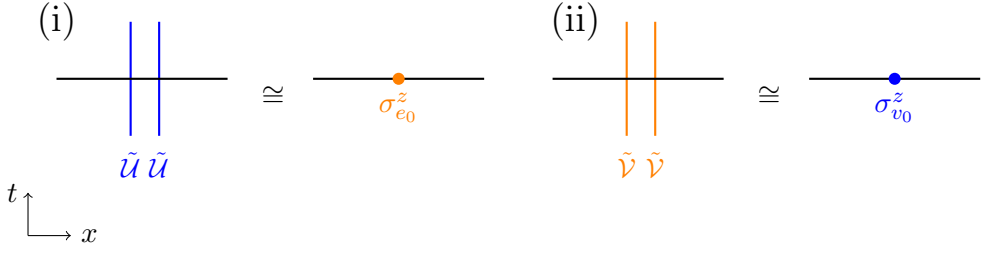}
    \caption{Fusing two symmetry defects results in an operator that is charged under the other symmetry, a consequence of the mixed anomaly.} \label{fig:SymFracDefects}
\end{figure}

In Sec.~\ref{subsec:subgauging}, we have reviewed that both $\Z_2$ symmetries fractionalize in the presence of a symmetry twist of the other symmetry, as a consequence of the mixed anomaly~\cite{Moradi2025}. Here, inspired by category theoretic arguments~\cite{Bhardwaj2023}, we propose two new ways of detecting symmetry fractionalization from a defect perspective: (1) two $\Z_2$ symmetry operators on a condensation defect of the other symmetry do not square to one but instead leave behind a residual $\Z_2$-symmetry operator, as depicted in Fig.~\ref{fig:SymFracOnCond}; (2) fusing two defects leaves behind an operator that is charged under the other symmetry, as depicted in Fig.~\ref{fig:SymFracDefects}. 

\paragraph{Fractionalization on a non-invertible condensation defect.}

Consider the setup specified in Fig.~\ref{fig:SymFracOnCond}. We fuse the two symmetry operators $\widetilde{U}$ of the $\Z_2^{(0)}\times_\epsilon\Z_2^{(d-1)}$-theory in the presence of a condensation defect of the $\Z_2^{(d-1)}$-symmetry. In \cite{Bhardwaj2023}, it is argued using categorical arguments that such a fusion leaves behind an operator localized on the condensation defect. 

\medskip \noindent Let us show this explicitly on the lattice. We localize the condensation defect on some $k$-dimensional sublattice $\Sigma_{k}$, where $k\leq d$. In other words, we gauge the $\Z_2^{(d-1)}$-symmetry, generated by $\widetilde{\cV}$ on $\Sigma_{k}$. We find
\begin{equation}
    \widetilde U = \prod_v \widetilde u_v \quad \text{where} \quad \widetilde u_v =\begin{cases}\sigma_v^x(P_v^++P_v^-A_v)\,\text{if } v\notin \Sigma_{k}\,,\\
        \sigma_v^x(P_v^++P_v^-\prod_{e\ni v,e\notin\Sigma_{k}}\sigma_e^z \tau_v^x) \, \text{if } v\in\Sigma_{k}
    \end{cases} \,.
\end{equation}
Here, $\tau_x$ acts on the auxiliary Hilbert space $\cH_{aux} = \C^2_v$. Note that this is a different Hilbert space than the original vertex Hilbert space that $\sigma_v^{x/z}$ acts on. A straightforward computation shows
\begin{equation}
    \widetilde U^2 = \prod_v A_v \prod_{v\in\Sigma_{k}}\left(\tau_v^x\prod_{e\ni v,e\in\Sigma_{k}}\sigma_e^z\right) = \prod_{v\in\Sigma_{k}}\tau_v^x
\end{equation}
where in the second equality we assumed that there are no symmetry twists and that $\Sigma_{k}$ is closed. We observe that $\cU$ does not square to one on a condensation defect. This can be understood using the isomorphism used for the sequential gauging picture in Sec.~\ref{subsec:SequentialGauging}. Even though $\cU$ is formally a $\Z_2^{(0)}$-symmetry operator, it realizes a $\Z_4^{(0)}$-symmetry on the condensation defect (on which the symmetry is gauged). $\cU_{v\in\Sigma_k}$ then corresponds to the $\Z_4$-symmetry operator by the isomorphism $\bigotimes_{v\in\Sigma_k}(\C^2\otimes \C^2)_v\cong \bigotimes_{v\in\Sigma_k}\C^4_v$ and $\cU^2= \prod_{v\in\Sigma_{k}}\tau_v^x$ describes its $\Z^{(0)}_2$ subgroup.

\medskip\noindent Let us now do the same for $\widetilde{\cV}$ on a $\Z_2^{(0)}$ condensation defect. From the above discussion, it follows that $\widetilde v_e$ acts as a $\Z_4$ operator $\widetilde{X}_e$ whenever gauged. Therefore, on a condensation defect, we find
\begin{equation}
    \widetilde V = \prod_{e\in\Sigma_1} \widetilde v_e \quad \text{where} \quad \widetilde{v}_e = \begin{cases}
        \frac{1}{2}(1-i\sigma^z_{s(e)})\sigma_e^x(1+i\sigma^z_{t(e)}), \quad \text{if } e \notin \Sigma_k, \\
        \widetilde{X}_e, \quad \text{if } e \in \Sigma_k.
    \end{cases}
\end{equation}
Therefore, we find
\begin{equation}
     \widetilde V^2 = \prod_{e\in\Sigma_k}\widetilde{X}_e^2\,,
\end{equation}
even without other symmetry twists. This confirms that the symmetry is fractionalized on a condensation defect. This provides an alternative route to detect symmetry fractionalization, without the need to resort to studying twisted sectors of the symmetry.

\paragraph{Fractionalization of invertible symmetry defects.}

Instead of looking at the symmetry operators, we can also consider fusing two symmetry defects, using the defect Hamiltonians as specified in~\eqref{eq:SubSymDefectU} and \eqref{eq:SubSymDefectV}. We will denote the defects of $\widetilde U$ and $\widetilde{V}$ with $\widetilde{\cU}$ and $\widetilde{\cV}$, respectively. Since they are both $\Z_2$-defects, one might naively assume that they square to one. Instead, we find that they fuse to a local operator that is charged under the other symmetry. This is illustrated in Fig.~\ref{fig:SymFracDefects}.
For simplicity, let us assume $d=1$ in the following. Though, we note that this calculation straightforwardly generalizes to higher dimensions, using the techniques developed in previous sections.

\medskip\noindent The Hamiltonian with two $\widetilde \cU$ defects, defined on edges $e_0$ and $e_0+1$ reads
\begin{align}
    \begin{split}
    H_{\widetilde{\cU}(e_0),\widetilde{\cU}(e_0+1)} &= -\sum_v \sigma_v^x(P_v^++P_v^-A_v) - \frac{1}{2}\sum_{e\neq e_0-1,e_0}(1-i\sigma_{s(e)}^z)\sigma_e^x(1+i\sigma_{t(e)}^z)\\& -\frac{i}{2}(1-i\sigma_{s(e_0)}^z)\sigma_{e_0}^x(1+i\sigma_{t(e_0)}^z)-\frac{i}{2}(1-i\sigma_{s(e_0+1)}^z)\sigma_{e_0+1}^x(1+i\sigma_{t(e_0+1)}^z)
    \end{split}
\end{align}
We can fuse the two defects by applying the movement operator $\widetilde u_{t(e_0-1)} = \widetilde u_{s(e_0)}$. This results in the Hamiltonian
\begin{align}
    \begin{split}
        H_{\widetilde{\cU}(e_0),\widetilde{\cU}(e_0+1)} &\cong -\sum_v \sigma_v^x(P_v^++P_v^-A_v) - \frac{1}{2}\sum_{e\neq e_0}(1-i\sigma_{s(e)}^z)\sigma_e^x(1+i\sigma_{t(e)}^z)\\ &\quad - \frac{1}{2}(1-i\sigma_{s(e_0)}^z)(-\sigma_{e_0}^x)(1+i\sigma_{t(e_0)}^z) + \text{h.c} \\
        &= \sigma_{e_0}^zH\sigma_{e_0}^z\,.
    \end{split}
\end{align}
We see that fusing two symmetry defects produces a local operator $\sigma_{e_0}^z$ at the location of the defect. This operator is uncharged under the $\widetilde U$ symmetry, but it has charge $-1$ under the $\widetilde V$ symmetry.

\medskip\noindent
Let us also show that a similar observation can be made for the $\widetilde V$-symmetry. Again, we start from a Hamiltonian with two defect insertions:
\begin{align}
    \begin{split}
    H_{\widetilde{\cV}(v_0),\widetilde{\cV}(v_0+1)} &= -\sum_{v\neq v_0,v_0+1}\sigma_v^x(P_v^++P_v^-A_v) - \frac{1}{2}\sum_{e}(1-i\sigma_{s(e)}^z)\sigma_e^x(1+i\sigma_{t(e)}^z)\\ &\quad- i\sigma_{v_0}^z\sigma_{v_0}^x(P_{v_0}^+-P_{v_0}^-A_{v_0}) - i\sigma_{v_0+1}^z\sigma_{v_0+1}^x(P_{v_0+1}^+-P_{v_0+1}^-A_{v_0+1})
    \end{split}
\end{align}
Now we apply the movement unitary $v_{e_0}$ with $e_0=(v_0,v_0+1)$ to find
\begin{align}
    \begin{split}
        H_{\widetilde{\cV}(v_0),\widetilde{\cV}(v_0+1)} &\cong -\sum_{v\neq v_0}\sigma_v^x(P_v^++P_v^-A_v) - \frac{1}{2}\sum_{e}(1-i\sigma_{s(e)}^z)\sigma_e^x(1+i\sigma_{t(e)}^z) \\
        &\quad + \sigma_{v_0}^x(P_{v_0}^+-P_{v_0}^-A_{v_0}) \\
        &= \sigma_{v_0}^zH\sigma_{v_0}^z\,.
    \end{split}
\end{align}
To summarize, in this subsection we have outlined two methods of diagnosing symmetry fractionalization that do not require access to the symmetry twisted sector.

\section{Conclusion}
\label{sec:conclusion}

In this paper, we have presented a comprehensive study of gauging interfaces and non-invertible defects on the lattice in higher dimensions. In particular, we explicitly constructed the interface Hamiltonians for $\mathbb Z_2^{(0)}$ gauging interfaces in $(2+1)d$ and for $\mathbb Z_2^{(1)}$ gauging interfaces in $(3+1)d$. In higher dimensions and in the presence of higher-form symmetries, the topological nature of gauging interfaces is obscured by the fact that the constrained Hilbert space depends on the location of the interface. We resolved this subtlety by constructing the appropriate movement operators, which transform the interface Hamiltonians and the constraints simultaneously.

\medskip\noindent Building on this construction, we further provided a detailed analysis of non-invertible condensation defects, the fusion between interfaces and defects, and the relations among different condensation defects under movement and measurement. In particular, we showed that the gauging map can be obtained from a decomposable condensation defect by sweeping the gauging interface across the entire spatial lattice. This also clarifies the connection to the sequential circuit representation of non-invertible duality transformations~\cite{Tantivasadakarn:2023zov,Tantivasadakarn:2025txn}.

\medskip\noindent 
Our construction can be straightforwardly generalized to arbitrary Abelian-group gauging interfaces in arbitrary spacetime dimensions. We also applied our framework to $\mathbb Z_4^{(0)}$ symmetry, for which gauging a subgroup is possible. By studying the gauging interfaces and defects associated with a $\mathbb Z_2^{(0)}$ subgroup, we showed the emergence of symmetry fractionalization from a defect perspective. This provides a new approach to diagnosing possible mixed anomalies using defects.

\medskip\noindent
There are several interesting directions for future work:
\begin{enumerate}
    \item \textbf{Extension to Hopf-algebra-valued qudits.} Recently, the self-duality of Abelian groups under gauging has been generalized to non-anomalous Rep$(H)$ symmetry~\cite{Choi:2023vgk,Lu:2025gpt}, where $H$ is a finite-dimensional semisimple Hopf algebra. In particular, generalized Ising models and generalized Kramers--Wannier dualities have been studied on tensor-product Hilbert spaces with Hopf-algebra-valued qudits~\cite{Lu:2026rhb}. It is therefore natural and interesting to extend our construction of gauging interfaces and defects to this more general setting.
    \item \textbf{Extension to modulated symmetries.} For modulated symmetries, such as dipole symmetries, subsystem symmetries, and fractal symmetries, the topological nature of invertible symmetry operators may be lost. It is therefore important to understand the topological behavior of gauging interfaces and defects for these more exotic symmetries. In $(2+1)d$, the defect associated with subsystem Kramers--Wannier symmetry has already been constructed~\cite{Cao:2023doz}. However, a complete understanding of the movement of such defects is still lacking. Extending our framework to these settings would therefore be highly interesting.
    \item \textbf{Construction of non-invertible defects.} Gauging interfaces are elementary building blocks for the construction of non-invertible defects. We have analyzed indecomposable condensation defects obtained as the product of a gauging interface and its conjugate along a non-contractible cycle. Our construction can be applied to defects associated with a broad class of non-invertible symmetries, namely double-coset symmetries~\cite{Hsin:2024aqb,Cao:2025qnc,Hsin:2025ria}, in which the non-invertible symmetry takes the form of a sandwich of the gauging map, a unitary operator, and the conjugate gauging map. We leave this for future study. 
    \item \textbf{Detection of LSM-type constraints.} LSM-type constraints can be understood as mixed 't Hooft anomaly between internal symmetry and lattice translation. It has recently attracted increasing interest because modulated symmetry~\cite{Ebisu:2025mtb} and non-invertible crystalline symmetry~\cite{Oishi:2026sow,Oishi:2026cvg} can arise from gauging internal symmetry with a LSM type anomaly in general dimensions. Besides traditional methods to detect such constraints, recently, new methods through gauging~\cite{Aksoy:2023hve} and defects~\cite{Yao:2020xcm,Yao:2023bnj,Seifnashri:2023dpa} have been proposed in $(1+1)d$, as well as new LSM constraints with generalized symmetry~\cite{Pace:2024acq,Pace:2025hpb,Furukawa:2025flp,Ning:2026noc}. It might be interesting to extend our framework and anomaly diagnosis for LSM-type anomalies in higher dimensions.
\end{enumerate}

\section*{Acknowledgments}
We thank Apoorv Tiwari for illuminating discussions and helpful comments on the draft. W.C. also thank the discussion with Linhao Li. D.H. and W.C. are funded by Villum Fonden Grant no. VIL60714. G.P. was financed in part by the Coordenação de Aperfeiçoamento de Pessoal de Nível Superior - (CAPES).

\appendix

\section{Details on the Movement Operators}
\label{app:Movement}
\subsection{All Elementary Two-Dimensional Moves}
Here, we verify explicitly that the movement operator presented in \ref{subsec:0formmovement} gives the correct results. For convenience, let us recall that the 2d movement operator is of the form
\begin{equation}
    \lambda_{e_0,v_0}^{0,1} = \left(\prod_{e\ni v_0,e\neq e_0} \mathsf{CZ}_{e,v_0}\right)\mathsf{H}_{v_0}\mathsf{S}_{e_0,v_0}\,.
\end{equation}
\begin{figure}
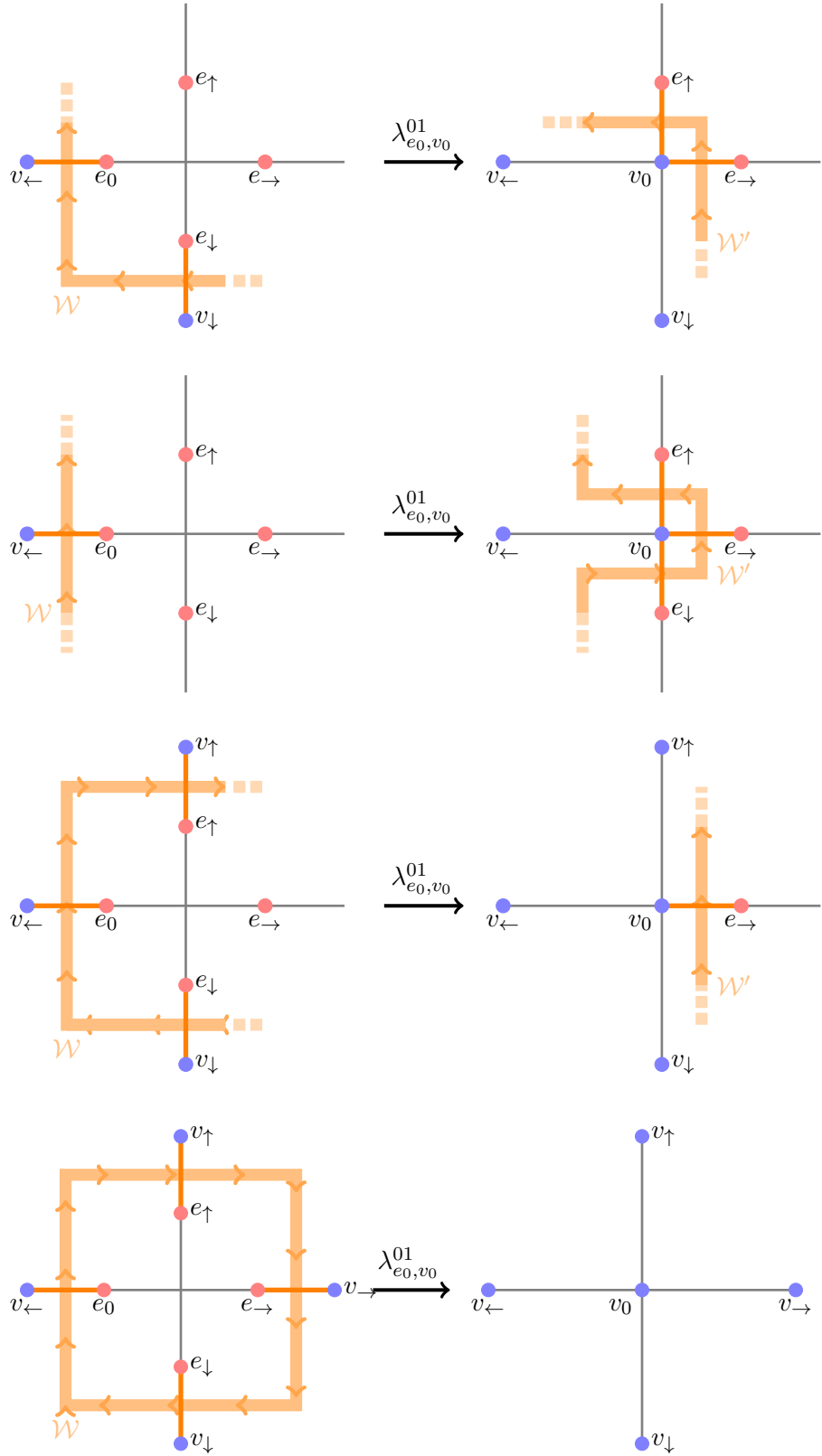

    \centering
    \begin{subfigure}{.8\textwidth}
        \includegraphics[width=\linewidth]{Figures/tikz/Mvmnt1.tikz}
        \phantomsubcaption
        \label{fig:2dMovement1}
    \end{subfigure}
    \begin{subfigure}{.8\textwidth}
        \includegraphics[width=\linewidth]{Figures/tikz/Mvmnt2.tikz}
        \phantomsubcaption
        \label{fig:2dMovement2}
    \end{subfigure}
    \begin{subfigure}{.8\textwidth}
        \includegraphics[width=\linewidth]{Figures/tikz/Mvmnt3.tikz}
        \phantomsubcaption
        \label{fig:2dMovement3}
    \end{subfigure}
    \begin{subfigure}{.8\textwidth}
        \includegraphics[width=\linewidth]{Figures/tikz/Mvmnt4.tikz}
        \phantomsubcaption
        \label{fig:2dMovement4}
    \end{subfigure}
    \caption{Up to rotations, there are four possible local movements in 2d. These are realized by the unitary $\lambda^{01}_{e_0,v_0}$ which removes an edge $e_0$ and adds the vertex $v_0$. Additionally, there may be constraints that trivialize after the movement, possibly projecting out other edges.}
    \label{fig:2dMovements}
\end{figure}
Let us now explicitly work out how this operator acts on the configurations presented in Figure~\ref{fig:2dMovements}.
\begin{enumerate}[label=(\alph*)]
    \item For the configuration in Figure~\ref{fig:2dMovement1}. The movement operator acts non-trivially on the following terms of the Hamiltonian
\begin{subequations}
   \begin{align}
       Z_{v_\leftarrow}X_{e_0} &\mapsto Z_{v_\leftarrow}Z_{v_0}\,, \\
       Z_{v_\downarrow}X_{e_\downarrow} &\mapsto Z_{v_\downarrow}X_{e_\downarrow}Z_{v_0}\,,\label{eq:2dExtraDof}\\
       A_{v_0}&\mapsto X_{v_0}\,,\\
       X_{e_\uparrow}&\mapsto Z_{v_0}X_{e_\uparrow}\,, \\
       X_{e_\rightarrow}&\mapsto Z_{v_0}X_{e_\rightarrow}\,.
   \end{align} 
\end{subequations}
It also changes the constraints
\begin{subequations}
    \begin{align}
        1= Z_{v_0} &\mapsto X_{e_0} = 1\,, \\
        1 = X_{e_0}X_{e_\downarrow} &\mapsto X_{e_\downarrow} = 1\,,
    \end{align}
\end{subequations}
where the second constraint comes from considering $V_{\gamma_0}$ with $\gamma_0$ being the degrees of freedom around the plaquette $p_0$.

Therefore, we indeed find
\begin{equation}
    \lambda^{0,1}_{e_0,v_0} H_{\cD(\cW)}P(\cW)\left(\lambda^{0,1}_{e_0,v_0}\right)^\dagger = H_{\cD(\cW')}P(\cW')\,.
\end{equation}

The Hamiltonian is now non-trivial on $v_0$ but does not act on $e_0$ and $e_\downarrow$ anymore. This may at first look like we have removed a degree of freedom (at odds with the unitarity of $\lambda^{0,1}_{e_0,v_0}$). However, the constraints enforce that we have $X_{e_0} = X_{e_\downarrow}$ before moving, so $X_{e_0}$ and $X_{e_\downarrow}$ actually represent a single degree of freedom. Therefore, the movement operator does not change the total number of degrees of freedom in the constrained Hilbert space, consistent with unitarity.
    \item For the configuration in Figure~\ref{fig:2dMovement2}, we have the following terms before the movement:
    \begin{align}
        \begin{split}
            Z_{v_\leftarrow} X_{e_0} &\mapsto Z_{v_\downarrow}Z_{v_0}\,,\\
            A_{v_0} &\mapsto X_{v_0}\,, \\
            X_{e} &\mapsto X_{e}Z_{v_0} \quad \text{for} \quad e\ni v_0, e\neq e_0\,,\\
            X_{v_0} &\mapsto X_{e_0}\,.
        \end{split}
    \end{align}
    The first three terms are exactly the changes we ask for in the Hamiltonian and the last term shows that after movement, the state on $X_{e_0}$ is trivialized.
    \item For Figure~\ref{fig:2dMovement3}, we have to consider the following terms:
    \begin{align}
        \begin{split}
            Z_{v_\leftarrow}X_{e_0} &\mapsto Z_{v_\leftarrow}Z_{v_0}\,, \\
            Z_{v_\downarrow}X_{e_\downarrow} &\mapsto Z_{v_\downarrow}X_{e\downarrow}Z_{v_0}\,, \\
            Z_{v_\uparrow}X_{e_\uparrow} &\mapsto Z_{v_\uparrow}X_{e\uparrow}Z_{v_0}\,, \\
            A_{v_0} &\mapsto X_{v_0}\,, \\
            X_{e_\leftarrow} &\mapsto Z_{v_0}X_{e_\leftarrow}\,.
        \end{split}
    \end{align}
    and for appropriate choice of loop $\gamma$, we find the additional constraints (see Eq.~\eqref{eq:2dConstraintsDualityDefect})
    \begin{align}
        \begin{split}
            1 = X_{e_0}X_{e_\uparrow} &\mapsto X_{e_\uparrow} = 1\,,\\
            1 = X_{e_0}X_{e_\downarrow} &\mapsto X_{e_\downarrow} = 1\,.
        \end{split}
    \end{align}
    \item The last movement, specified in Figure~\ref{fig:2dMovement4} is an interesting special case as it relates to the discussion in Section~\ref{sec:0formCond} on contractible condensation defects. We have
    \begin{align}
        \begin{split}
            Z_{v_\leftarrow}X_{e_0} &\mapsto Z_{v_\leftarrow}Z_{v_0}\,, \\
            Z_{v_\uparrow}X_{e_\uparrow} &\mapsto Z_{v_\uparrow}X_{e_\uparrow}Z_{v_0} \,,\\
            Z_{v_\downarrow}X_{e_\downarrow} &\mapsto Z_{v_\downarrow}X_{e_\downarrow}Z_{v_0}\,, \\
            Z_{v_\rightarrow}X_{e_\rightarrow} &\mapsto Z_{v_\rightarrow}X_{e_\rightarrow}Z_{v_0}\,, \\
            A_{v_0} &\mapsto X_{v_0}\,.
        \end{split}
    \end{align}
    We have the constraints
    \begin{align}
        \begin{split}
        1 = X_{e_0}X_{e_\uparrow} &\mapsto X_{e_\uparrow} = 1\,, \\
        1 = X_{e_0}X_{e_\downarrow} &\mapsto X_{e_\downarrow} = 1\,, \\
        1 = X_{e_\uparrow}X_{e_\rightarrow} &\mapsto  X_{e_\uparrow}X_{e_\rightarrow} = 1\,.
        \end{split}
    \end{align}
    Together, they imply that all edges trivialize. Since any contractible condensation defect can be brought into this form by the other three movements, this gives another perspective on the fact that all contractible condensation defects of 0-form symmetry are trivial in 2d.
\end{enumerate}

\subsection{Gauging Map From Sweeping the Interface}
Here, we go explicitly through the sweeping of the gauging interface to show that the procedure outlined in Sections~\ref{Subsec:RelatingCondDefects} and~\ref{subsec:2dOperatorFromDefect} indeed produces the correct gauging map. For convenience, we recall the relevant figures here, in Figures~\ref{fig:Appendix1dMovementCondensationDefect} and~\ref{fig:AppendixDualityFromMovement}.
\begin{figure}
    \centering
    \begin{subfigure}{\textwidth}
    \includegraphics[width=\linewidth]{Figures/tikz/1dMovementCondensationDefect.tikz}
    \caption{After successive application of the movement operator, we can relate the decomposable defect $C_1(\Sigma_0)$ to the indecomposable $C_1(\Sigma_1)$.}
    \label{fig:Appendix1dMovementCondensationDefect}
    \end{subfigure}
    \begin{subfigure}{\textwidth}
    \includegraphics[width=\linewidth]{Figures/tikz/OperatorFromCondDefect.tikz}
    \caption{Starting from a one-dimensional condensation defect $\cC_1(\Sigma_1)$ in the $\Z_2^{(1)}$-theory (and its twisted sector), we use the unitary movement operator to progressively enlarge it until the gauged region fills the entire lattice.}
    \label{fig:AppendixDualityFromMovement}
    \end{subfigure}
    \phantomcaption
\end{figure}

We begin with the $\Z_2^{(1)}$-theory with a single extra degree of freedom on $v_0 = \Sigma_0$ that represents a 0-dimensional, decomposable condensation defect $\cC_1(\Sigma_0) = 1\oplus\eta_1(\Sigma_0)$ on $\Sigma_0$. Then, using the procedure outlined in Section~\ref{Subsec:RelatingCondDefects}, we can extend this to a one-dimensional condensation defect on $\Sigma_1$ with a $\Z_2^{(0)}$-defect $\widetilde{\eta}_0$ living on the condensation defect. The operator that achieves this is of the form.
\begin{equation}
    \mathsf{C}_1^{1\oplus \widetilde\eta_0}(\Sigma_1) = \left( \prod_{i\in{1,...,n}}^\leftarrow\lambda^{0,0}_{e_{i},v_i}\right) \mathsf{CZ}_{v_0,e_0}\mathsf{CZ}_{v_0,e_1}\ket{+}_{v_0}\,,
    \label{eq:App2dCondOperator}
\end{equation}
In the main text, we have shown that applying these unitaries to the defect Hamiltonian produces the correct condensation defect. In preparation of the ensuing discussion, let us show the analogous results for the operators themselves. We find:
\begin{subequations}
    \begin{align}
        \mathsf{C}^{1\oplus\widetilde{\eta}}_1(\Sigma_1): A_{v_0}X_{v_0}&\mapsto X_{v_0} \\ 
        A_{v_i} &\mapsto X_{v_i} \qquad \qquad \text{for } v_i\neq v_0 \\
        X_{e_0} &\mapsto Z_{v_n}X_{e_0}Z_{v_0} \\
        X_{e_i} &\mapsto Z_{v_{i-1}}Z_{v_i} \qquad \text{for } e_i\neq e_0
    \end{align}
\end{subequations}
This can be seen as the first step of sweeping the 0-dimensional $\Z_2^{(1)}$-defect across the lattice to realize the gauging. Indeed, there is a $\Z_2^{(0)}$-symmetry on the condensation defect on $\Sigma_1$.

As a next step, we apply $\mathsf{CZ}$ gates of the form
\[
    \prod_{v\in \Sigma_1}\prod_{e\ni v,e\notin \Sigma_1}\mathsf{CZ}_{v,e}
\]
to make the condensation defect admissible for application of the 2d movement operator. \footnote{Alternatively, these gates may be absorbed into the movement operators in the definition of the condensation defect creation operator in Eq. \eqref{eq:App2dCondOperator}, where they have the effect of changing $\lambda^{0,0}_{e,v}$ into $\lambda^{0,1}_{e,v}$.} Next, we now further apply movement operators along the direction orthogonal to $\Sigma_1$. 
This enlarges the defect $\widetilde\eta_0(e_0) \to \eta_0({\Sigma_1^\vee}^{\perp})$ where $\Sigma_1^{\perp}$ is the non-contractible cycle orthogonal to $\Sigma_1$. In the last step, this procedure creates another defect $\eta_0(\Sigma_1^\vee)$ by fusing the two duality interfaces according the usual $\cD\otimes\cD^\dagger \cong 1\oplus \eta_0$ fusion rule. This is sketched in Figure\ref{fig:AppendixDualityFromMovement}. The full operator reads
\begin{equation}
    \mathsf{D}^{\dagger} = \bra{+}_{\Sigma_1^\vee}\bra{+}_{{\Sigma_1^\vee}^\perp}\left(\prod_{i=0,...,L-1}^{\leftarrow}\Lambda^{0,1}_{\cW^{(i)},\cW^{(i+1)}}\right)\left(\prod_{v\in \Sigma_1}\prod_{e\ni v,e\notin \Sigma_1}\mathsf{CZ}_{v,e}\right)\mathsf{C}_1(\Sigma_1)^{1\oplus \widetilde\eta_0}
    \label{eq:AppOpFromDefect2d}
\end{equation}
Let us check very explicitly that defining gauging this way realizes the correct mapping 
\begin{equation}
    \mathsf{D}^\dagger:\{A_v\,, X_e\,, \prod_{e\in \gamma}X_e\}\longmapsto\{X_v\,, Z_{s(e)}Z_{t(e)}\,, V_{\gamma}\} \,.
\end{equation}
where $V_\gamma = \prod_{e\in \gamma} Z_{s(e)}Z_{t(e)}$ measures the $\mathbb{Z}_2^{(0)}$ symmetry twist. We have already shown this for any operator on $\Sigma_1$. Therefore, let us now take arbitrary operators localized somewhere in the bulk. 
The unitary part acts as 
\begin{subequations}
    \begin{align}
        A_v &\mapsto X_v \,,\\
        X_e &\mapsto Z_{s(e)}Z_{t(e)} \qquad \text{for } e\notin \Sigma_1^\perp \text{ a horizontal edge}\,\,,\\
        X_e &\mapsto Z_{s(e)}X_eZ_{t(e)} \qquad \text{for } e \text{ a vertical edge or }e\in \Sigma_1^\perp \,.
    \end{align}
\end{subequations}
We see that at this point, we have successfully transformed the horizontal edges into the correct vertex degrees of freedom. However, the vertical edges seemingly remain. Similarly to the elementary movements considered above, we can resolve this issue by tracking how the topological symmetry constraint transforms after sweeping. Consider a minimal symmetry operator $\prod_{e\in p}X_e$. Its transformation depends on its location in the lattice. Generally, we have
\begin{equation}
    1=\tikzXXXX \mapsto \tikzXXhor =1
    \label{eq:TopSymAfterSweep}
\end{equation}
There are two cases, in which the constraint transforms differently: First, when $p\in \Sigma_1^\vee$, we have 
\begin{equation}
    1=\tikzXXXX\mapsto \tikzXXX
\end{equation}
If the plaquette is just to the right of the boundary, we will instead get
\begin{equation}
    1=\tikzXXXX \mapsto \tikzXsquare\,,
\end{equation}
except for the case when $e_0\in p$, when we still have that Eq.~\eqref{eq:TopSymAfterSweep} holds. Together with Eq.~\eqref{eq:TopSymAfterSweep}, we can view these constraints on $\Sigma_1$ as "seeds" that enforce $X_e = 1$ on vertical bonds throughout most of the lattice. On the other hand, when $e\in \Sigma_1^\perp$, we do not have this seed and we can not resolve this product. Instead, there are now two ways of resolving the constraints: Either $X_e = 1$, or $X_e=-1$ for all ${e\in\Sigma_1^\vee}^\perp$. The same holds true when $X_e\in \Sigma_1^\vee$. 
After explicitly projecting onto the trivial symmetry sector, we have arrived at the gauged theory with dual $\Z_2^{(0)}$-symmetry. 

\paragraph{Comparison to other dimensions.}
Let us compare this to the 1d case, where the Kramers-Wannier operator can be written as inserting a decomposable, point-like condensation defect of the 0-form symmetry, then moving one of the interfaces around the lattice, and finally fusing them again and projecting onto the trivial fusion channel \cite{Okada2024,Seiberg2024}. Equation~\eqref{eq:AppOpFromDefect2d} is the analogous operator in $(2+1)d$, implementing the duality between the $\Z_2$ gauge theory and the transverse-field Ising model by only inserting a condensation defect at a single site. To summarize, both in 1d and 2d, the gauging operator for $\Z_2^{(0)}$ symmetry can be constructed in the following steps:
\begin{enumerate}
    \item Insert a decomposable condensation defect.
    \item Use unitaries to move the duality interface, making up the boundary of the condensation defect, around the whole lattice.
    \item Project out the resulting non-trivial defects in the dual theory. (These are again decomposable into simply symmetry defects.)
\end{enumerate}
We expect that this story remains true in arbitrary dimensions. For example, for three spatial dimensions, one would have to insert a dual, decomposable one-dimensional $\Z_2^{(1)}$-defect $1\oplus \eta_1(\Sigma_1)$. One would then use the movement operator constructed in Section~\ref{subsec:1formmovement} to relate this to a two-dimensional indecomposable condensation defect, which in turn can further be expanded into the dual $\Z_2^{(1)}$-theory with dual symmetry defects that would need to be projected out.

\bibliography{biblio.bib}

\end{document}